\documentclass[twocolumn,aps,pra,amsmath,amssymb,notitlepage]{revtex4-1}
\usepackage[latin9]{inputenc}
\usepackage{amsmath}
\usepackage{amssymb}
\usepackage{graphicx}
\usepackage{esint}

\makeatletter

\newcommand*\LyXThinSpace{\,\hspace{0pt}}

\usepackage{epsfig}
\usepackage{bm}
\usepackage{dcolumn}
\usepackage{color}

\makeatother

\begin{document}
\title{Collective excitations of a spherical ultradilute quantum droplet}
\author{Hui Hu and Xia-Ji Liu}
\affiliation{Centre for Quantum Technology Theory, Swinburne University of Technology,
Melbourne, Victoria 3122, Australia}
\date{\today}
\begin{abstract}
In three dimensions, exotic new state of matter of self-bound ultradilute
quantum droplets can be realized in free space, when the mean-field
attraction (i.e., with mean-field energy $E_{\textrm{MF}}\propto-n^{2}$
at the density $n$) is balanced by the repulsive beyond-mean-field
quantum fluctuations (i.e., $E_{\textrm{BMF}}\propto n^{2+\gamma}$).
The parameter $\gamma>0$ typically takes the value $1/2$ if we consider
the Lee-Huang-Yang (LHY) energy functional, but it can vary when the
beyond-LHY-effect becomes important or the three-body interaction
becomes dominant. Here, we theoretically investigate how collective
excitations of a three-dimensional quantum droplet are affected by
the parameter $\gamma$ and a weak harmonic trapping potential, both
of which could be tuned in experiments. We use both the approximate
approach based on a Gaussian variational ansatz and the exact numerical
solution of the Bogoliubov equations resulting from the linearized
time-dependent extended Gross-Pitaevskii equation. We show that one
of the key features of quantum droplets, i.e., the existence of the
surface modes with dispersion relation $\omega_{s}\propto k^{3/2}$
is very robust with respect to the changes either in the parameter
$\gamma$ or in the harmonic trapping potential. We predict the excitation
spectrum of the droplet realized by binary $^{39}$K mixtures under
the typical experimental conditions, which might be readily measured
in current cold-atom laboratories. 
\end{abstract}
\maketitle

\section{Introduction}

Over the past five years, the theoretical proposal \citep{Petrov2015}
and the experimental realization of an ultradilute quantum droplet
with cold-atoms \citep{FerrierBarbut2016,Schmitt2016,Chomaz2016,Cabrera2018,Cheiney2018,Semeghini2018,Ferioli2019,Tanzi2019PRL,DErrico2019,Bottcher2019}
open a new paradigm to investigate the intriguing quantum many-body
physics \citep{Bottcher2020}. This new state of matter builds on
the delicate balance between the mean-field attraction and the repulsive
force resulting from beyond-mean-field quantum fluctuations \citep{Petrov2015}.
At the leading order, the energy functional responsible for the repulsive
force was worked out by Lee, Huang and Yang (LHY) in their seminal
work long time ago \citep{LeeHuangYang1957} and takes the form $E_{\textrm{LHY}}\propto n^{2+\gamma}$
with the parameter $\gamma=1/2$ at the density $n$. To date, the
formation of quantum droplets has been observed both in single-component
Bose gases with long-range dipolar interactions \citep{FerrierBarbut2016,Schmitt2016,Chomaz2016,Tanzi2019PRL,Bottcher2019}
and in two-component or binary Bose mixtures with short-range inter-species
attractions \citep{Cabrera2018,Cheiney2018,Semeghini2018,Ferioli2019,Tanzi2019PRL,DErrico2019}.
The experimental observations can be qualitatively understood by using
an extended Gross-Pitaevskii equation (GPE) with the LHY energy functional
$E_{\textrm{LHY}}$ \citep{Bottcher2020}.

In this work, we aim to theoretically understand the collective excitations
of a three-dimensional \emph{spherical} ultradilute quantum droplet,
focusing on the experimental feasibility of observing the surface
modes with exotic dispersion relation $\omega_{s}\propto k^{3/2}$,
whose presence is one of the key features of quantum droplets \citep{Chin1995}.
The accurate frequency measurement of collective modes is known as
a powerful probe of the many-body state of ultracold quantum gases
\citep{Dalfovo1999}. For example, for strongly interacting Fermi
gases, the measurement of breathing modes provides the first indirect
proof of fermionic superfluidity in three dimensions \citep{Kinast2004,Bartenstein2004,Hu2004}
and quantum anomaly in two dimensions \citep{Holten2018,Peppler2018,Hu2019,Yin2020}.
For dipolar Bose gases, the most recent collective mode measurement
in arrays of dipolar droplets clearly shows the symmetry breaking
and the supersolid nature of the system \citep{Tanzi2019Nature,Guo2019}.
Therefore, it is natural to anticipate that the observation of the
surface modes in a three-dimensional spherical droplet would be an
excellent way to characterize this intriguing new state of matter
in on-going experiments. 

For simplicity, we follow the original proposal by Petrov \citep{Petrov2015}
and consider a binary Bose mixture with attractive inter-species interactions,
in which a spherical self-bound droplet has been observed \citep{Semeghini2018}.
In contrast, the dipolar quantum droplet is anisotropic and its cigar-shape
geometry disfavors the surface modes \citep{Baillie2016,Wachtler2016,Baillie2017}.
Actually, in Petrov's seminal proposal \citep{Petrov2015}, the excitation
spectrum $\omega_{l,n=0}$ of a spherical droplet, including the lowest
monopole mode (i.e., the breathing mode with $l=n=0$) and surface
modes ($l\geq2$ and $n=0$), has already been discussed, by taking
the LHY energy functional (i.e., $\gamma=1/2$). Here, $l$ is the
angular momentum of the modes and the non-negative integer $n$ (not
be confused with the density) characterizes the orders of the excitations
\citep{NoteNumberNodes}. The motivation of our collective excitation
study is two-fold.

First, we wish to understand how the whole excitation spectrum $\omega_{ln}$
(including the modes with $n\neq0$ not addressed by Petrov) changes
when the parameter $\gamma$ deviates from the idealized case of the
LHY exponent $\gamma=1/2$. In the experiments with binary bosonic
mixtures \citep{Cabrera2018,Semeghini2018}, for example, the $^{39}$K-$^{39}$K
mixture, the energy functional resulting from the beyond-mean-field
effects does not necessarily take the LHY form. As shown by recent
diffusion Monte Carlo (DMC) simulations \citep{Cikojevic2019,Cikojevic2020,Cikojevic2020arXiv},
the beyond-LHY effect and the finite-range correction to the short-range
interactions \emph{effectively} lead to a smaller parameter $\gamma$,
which can be as small as 0.276 under the realistic experimental conditions
\citep{Cikojevic2020arXiv}. On the other hand, if we consider the
possible (yet to be realized) quantum droplet formed by the three-body
repulsion in tritium condensates \citep{Bulgac2002,Blume2002,Mestrom2020},
the parameter $\gamma=1$. Interestingly, if we think more broadly
and include superfluid helium nano-droplets \citep{Stringari1987},
the parameter $\gamma$ can be as large as $2.8$.

Second, in the current experiments for cold-atom droplets, it is difficult
to fully eliminate the external harmonic trapping potential. For instance,
in the experiment performed at the European Laboratory for Non-linear
Spectroscopy (LENS), the \emph{residual} trapping frequency along
the axial direction is estimated to be $\omega_{z}=2\pi\times12$
Hz \citep{Semeghini2018}. Therefore, theoretically it would be important
to understand how the properties of a quantum droplet are affected
by this weak residual harmonic trapping potential. On the other hand,
the stability of the quantum droplet might be improved by keeping
a finite harmonic trapping potential. In this respect, the collective
excitations of the droplet in the presence of an external trapping
potential, particularly the existence of the surface modes, is an
interesting issue to consider in its own right.

Here, we would like to systematically investigate how collective excitations
are affected by the parameter $\gamma$ and by the finite external
trapping potential, based on the approximate variational approach
with a Gaussian ansatz and the exact numerical solution of the Bogoliubov
equations for density oscillations. We find that the peculiar surface
modes have weak dependences on the parameter $\gamma$ and on the
external trapping frequency and therefore should be able to manifest
themselves in future measurements. For their experimental observation,
we provide predictions for the excitation spectrum of the $^{39}$K
quantum droplet under the realistic experimental conditions.

We note that the breathing mode ($\omega_{00}$) and quadrupole mode
($\omega_{20}$) of a self-bound spherical $^{39}$K droplet are most
recently studied by Cikojevi\'{c} and his co-workers \citep{Cikojevic2020arXiv},
by using the time-dependent extended GPE equation \citep{Ferioli2020}
together with the accurate DMC energy functional, which takes an effective
exponent $\gamma<1/2$. Our work complement their studies by providing
the whole excitation spectrum and by accounting for the finite external
trapping potential. We note also that, collective excitations of a
quantum droplet in quasi-one-dimension \citep{Cappellaro2018} or
one-dimension \citep{Astrakharchik2018,Tylutki2020} have been recently
investigated by using both Gaussian variational approach and the Bogoliubov
equations. The approximate variational approach is shown to work well
in the limits of small and large particle numbers \citep{Astrakharchik2018}.
In this work, the validity of the approximate Gaussian variational
approach in three dimensions will be examined.

The rest of the paper is organized as follows. In the next section
(Sec. II), we introduce the extended GPE with suitable energy functional
as an effective low-energy description of quantum droplet states.
In Sec. III, we present the details of the approximate variational
approach with Gaussian ansatz and the numerical solutions of the Bogoliubov
equations derived from the linearized time-dependent extended GPE.
In Sec. IV, we discuss the properties of a three-dimensional spherical
quantum droplet in free space without harmonic trapping potential.
We show that the collective modes below the particle-emission threshold
can be well classified as the bulk modes and surface modes. The former
corresponds to the well-known sound modes (i.e., $\omega=ck$), while
the latter features the exotic dispersion relation $\omega_{s}\propto k^{3/2}$
at low momentum, once we properly define the wave-vector $k$ for
the discrete spectrum. In Sec. V, we consider quantum droplets in
the presence of a finite external trapping potential. We show that
the qualitative behavior of collective excitations does not change
under weak trapping potentials and the peculiar $k^{3/2}$ dispersion
persists. In Sec. VI, we make connection with the experiment, by calculating
the excitation spectrum of a $^{39}$K quantum droplet under the realistic
experimental conditions. Finally, Sec. VII is devoted to the conclusions
and outlooks.

\section{Time-dependent extended Gross-Pitaevskii theory}

We start from the extended GPE, which has been extensively used in
the past theoretical studies to describe the structure and dynamics
of quantum droplets \citep{Petrov2015,Bottcher2020},
\begin{equation}
i\hbar\frac{\partial\Phi}{\partial t}=\left[-\frac{\hbar^{2}}{2m}\nabla^{2}+\frac{m}{2}\omega_{T}^{2}\mathbf{x}^{2}-\mu_{a}+\frac{\partial E}{\partial n}\left(n=\left|\Phi\right|^{2}\right)\right]\Phi.\label{eq:EGPE}
\end{equation}
Here, $\Phi(\mathbf{x},t)$ can be treated as the condensate wave-function
of the droplet with mass $m$ and a chemical potential $\mu_{a}$
determined by the total number of particles $N_{a}$, and the total
energy functional $E(n)$ includes both the mean-field part $E_{\textrm{MF}}/V=-A_{0}n^{2}$
and the beyond-mean-field contribution $E_{\textrm{BMF}}/V=A_{1}n^{2+\gamma}$
($\gamma>0$). We also consider an external harmonic trapping potential
with frequency $\omega_{T}$, i.e., $m\omega_{T}^{2}\mathbf{x}^{2}/2$.
The extended GPE is often viewed as a \emph{phenomenological} low-energy
effective theory, following the seminal proposal by Petrov, where
the LHY energy functional $E_{\textrm{LHY}}\propto n^{5/2}$ is considered
\citep{Petrov2015}. In this work, we do not care about the microscopic
details of the theory and use Eq. (\ref{eq:EGPE}) in a broader context
to describe a general quantum droplet, created in the binary Bose
mixtures ($0<\gamma\leq1/2$) \citep{Cikojevic2019}, tritium condensates
($\gamma=1$) \citep{Mestrom2020}, and helium clusters ($\gamma\simeq2.8$)
\citep{Stringari1987}. 

Nevertheless, it is worth noting that the extended GPE can actually
be derived microscopically by applying a pairing theory to a two-component
Bose mixture with intra-species scattering length $a>0$ and inter-species
scattering length $a_{12}\sim-a<0$ \citep{Hu2020a,Hu2020b}. The
pairing is induced by the attractive inter-species interactions and
is robust in the case of equal spin-populations. The two components
are therefore perfectly locked together, with negligible spin-density
fluctuations at zero temperature. The low-energy collective excitations
described by Eq. (\ref{eq:EGPE}) correspond to the \emph{phase} fluctuations
of the pairing order parameter \citep{Hu2020c}, which have much lower
energy than the spin-density fluctuations. The latter are basically
the amplitude fluctuations of the pairing order parameter and have
a characteristic energy scale of the pairing gap \citep{Hu2020c}.
More quantitatively, the energy cost of the spin-density fluctuations
can be estimated to be about $\sqrt{a/\left|a+a_{12}\right|}$ times
larger than that of the collective excitations \citep{Petrov2015}.
Experimentally, we have $a_{12}\sim-1.05a$ \citep{Cabrera2018,Semeghini2018}
and consequently the ratio $\sqrt{a/\left|a+a_{12}\right|}\gg1$.
Thus, the spin-density fluctuations can hardly be excited at the typical
energy scale of collective excitations.

In the absence of the harmonic trapping potential ($\omega_{T}=0$)
and for a sufficiently large number of particles $N_{a}\gg1$, the
self-bound quantum droplet described by Eq. (\ref{eq:EGPE}) has an
equilibrium density $n_{0}$ in the bulk, which is set by the zero
pressure condition $P=(n\mu_{a}-E/V)_{n=n_{0}}=0$ at zero temperature:
\begin{equation}
n_{0}=\left[\frac{A_{0}}{\left(1+\gamma\right)A_{1}}\right]^{1/\gamma}.\label{eq:N0}
\end{equation}
Following Petrov \citep{Petrov2015}, it is convenient to define the
units of length $\xi,$ energy $\hbar^{2}/(m\xi^{2})$ and time $m\xi^{2}/\hbar$,
and introduce the re-scaled coordinate $\mathbf{r}=\mathbf{x}/\xi$,
time $\tau=\hbar t/(m\xi^{2})$, condensate wave-function $\phi=\Phi/\sqrt{n_{0}}$,
as well as the re-scaled frequency $\omega_{0}=\hbar\omega_{T}/[\hbar^{2}/(m\xi^{2})]$
and the chemical potential $\mu=\mu_{a}/[\hbar^{2}/(m\xi^{2})]$.
Therefore, we rewrite the extended GPE into a simpler dimensionless
form,
\begin{equation}
i\hbar\frac{\partial}{\partial\tau}\phi=\left[-\frac{1}{2}\nabla^{2}+\frac{1}{2}\omega_{0}^{2}r^{2}-\mu+\frac{\partial\epsilon}{\partial n}\left(\phi,\phi^{*}\right)\right]\phi,\label{eq:dimensionlessEGPE}
\end{equation}
where the dimensionless total energy functional and its derivative
are given by,
\begin{eqnarray}
\epsilon\left(\phi,\phi^{*}\right) & = & -\left(1+\gamma\right)\left|\phi\right|^{4}+\left|\phi\right|^{4+2\gamma},\\
\frac{\partial\epsilon}{\partial n}\left(\phi,\phi^{*}\right) & = & -2\left(1+\gamma\right)\left|\phi\right|^{2}+\left(2+\gamma\right)\left|\phi\right|^{2+2\gamma},
\end{eqnarray}
respectively. It is straightforward to show that the length scale
$\xi$ is determined by,
\begin{equation}
\frac{\hbar^{2}}{m\xi^{2}}=\frac{A_{0}n_{0}}{\left(1+\gamma\right)}=\left[\frac{A_{0}}{1+\gamma}\right]^{(1+\gamma)/\gamma}A_{1}^{-1/\gamma},\label{eq:EnergyUnits}
\end{equation}
which sets the energy scale of the droplet. The condensate wave-function
$\phi(\mathbf{r},\tau)$ should now be normalized according to,
\begin{equation}
\int d\mathbf{r}\left|\phi\left(\mathbf{r},\tau\right)\right|^{2}=\frac{N_{a}}{n_{0}\xi^{3}}\equiv N.
\end{equation}
In the absence of the harmonic trapping potential ($\omega_{0}=0$)
and in the thermodynamic limit ($N\rightarrow\infty$), the surface
effect of the droplet can be neglected and we have the uniform solution
$\phi^{(\infty)}=1$, with the chemical potential $\mu^{(\infty)}=-\gamma<0$.
We note that, with the LHY exponent $\gamma=1/2$ and without the
trapping potential $\omega_{0}=0$, the dimensionless extended GPE
Eq. (\ref{eq:dimensionlessEGPE}) has been solved by Petrov \citep{Petrov2015}.

\section{Gaussian variational approach and the Bogoliubov equations}

To solve the dimensionless extended GPE for a general parameter $\gamma$
and in the presence of the external harmonic trapping potential, we
use either the variational approach with a Gaussian ansatz or the
numerical solution of the Bogoliubov equations.

\subsection{Gaussian variational approach}

The Gaussian variational approach provides a useful qualitative description
of quantum droplets with the following simple normalized ansatz \citep{Cappellaro2018,Cappellaro2017},
\begin{equation}
\phi_{0}\left(r\right)=\frac{\sqrt{N}}{\pi^{3/4}\sigma^{3/2}}\exp\left[-\frac{r^{2}}{2\sigma^{2}}\right],
\end{equation}
where the subscript ``0'' indicates that the wave-function is time-independent,
and the width $\sigma$ is the only variational parameter, to be determined
by minimizing the (dimensionless) total energy,
\begin{equation}
\epsilon_{\textrm{tot}}=\int d\mathbf{r}\left[\frac{1}{2}\left(\nabla\phi_{0}\right)^{2}+\frac{1}{2}\omega_{0}^{2}r^{2}\phi_{0}^{2}+\epsilon\left(\phi_{0},\phi_{0}^{*}\right)\right].
\end{equation}
By substituting the ansatz into the total energy and performing the
integrals, it is easy to obtain,
\begin{equation}
\frac{\epsilon_{\textrm{tot}}}{N}=\frac{3}{4\sigma^{2}}+\frac{3}{4}\omega_{0}^{2}\sigma^{2}-\frac{\left(1+\gamma\right)N}{\left(2\pi\right)^{3/2}\sigma^{3}}+\frac{N^{1+\gamma}\sigma^{-3\left(1+\gamma\right)}}{\left(2+\gamma\right)^{3/2}\pi^{3\left(1+\gamma\right)/2}}.\label{eq:GaussianEtot}
\end{equation}
By minimizing the total energy, we find that,
\begin{equation}
\omega_{0}^{2}=\sigma^{-4}-\frac{2\left(1+\gamma\right)N}{\left(2\pi\right)^{3/2}\sigma^{5}}+\frac{2\left(1+\gamma\right)N^{1+\gamma}\sigma^{-5-3\gamma}}{\left(2+\gamma\right)^{3/2}\pi^{3\left(1+\gamma\right)/2}},\label{eq:GaussianWidth}
\end{equation}
from which, we numerically determine $\sigma$ for a given set of
parameters ($\gamma$, $\omega_{0}$, $N$). The above equation also
allows us to directly calculate the breathing mode frequency $\omega_{B}$,
by using the elegant sum-rule approach \citep{Hu2019,Menotti2002,Hu2014},
\begin{equation}
\omega_{B}^{2}=-2\frac{\left\langle r^{2}\right\rangle }{\partial\left\langle r^{2}\right\rangle /\partial\omega_{0}^{2}}=-\sigma\frac{\partial\omega_{0}^{2}}{\partial\sigma}.
\end{equation}
This leads to the expression,
\begin{equation}
\omega_{B}^{2}=\frac{4}{\sigma^{4}}-\frac{10\left(1+\gamma\right)N}{\left(2\pi\right)^{3/2}\sigma^{5}}+\frac{2\left(1+\gamma\right)\left(5+3\gamma\right)N^{1+\gamma}\sigma^{-5-3\gamma}}{\left(2+\gamma\right)^{3/2}\pi^{3\left(1+\gamma\right)/2}}.\label{eq:GaussianWB}
\end{equation}
For an ideal gas with $\epsilon(\phi,\phi^{*})=0$ (or $N=0$), from
Eq. (\ref{eq:GaussianWidth}) and Eq. (\ref{eq:GaussianWB}), we obtain
$\sigma=\omega_{0}^{-1/2}$ and $\omega_{B}=2\omega_{0}$, as one
may naively anticipate.

\subsection{Bogoliubov theory}

To quantitatively determine the ground-state profile and the collective
excitations of the quantum droplet, it is necessarily to solve the
stationary extended GPE for the condensate wave-function $\phi_{0}\geq0$,
\begin{equation}
\mathcal{\hat{L}}\phi_{0}\left(r\right)=\mu\phi_{0}\left(r\right),\label{eq:stationaryEGPE}
\end{equation}
and the Bogoliubov equations for small fluctuation modes around the
condensate (labeled by an integer $j$),
\begin{equation}
\left[\begin{array}{cc}
\mathcal{\hat{L}}-\mu+\mathcal{\hat{M}} & \mathcal{\hat{M}}\\
\mathcal{\hat{M}} & \mathcal{\hat{L}}-\mu+\mathcal{\hat{M}}
\end{array}\right]\left[\begin{array}{c}
u_{j}\left(\mathbf{r}\right)\\
v_{j}\left(\mathbf{r}\right)
\end{array}\right]=\omega_{j}\left[\begin{array}{c}
+u_{j}\left(\mathbf{r}\right)\\
-v_{j}\left(\mathbf{r}\right)
\end{array}\right],\label{eq:BogoliubovEQs}
\end{equation}
where we have defined the operators, 
\begin{eqnarray}
\mathcal{\hat{L}} & \equiv & -\frac{\nabla^{2}}{2}+\frac{\omega_{0}^{2}r^{2}}{2}+\frac{\partial\epsilon}{\partial n}\left(\phi_{0},\phi_{0}^{*}\right),\\
\mathcal{\hat{M}} & \equiv & -2\left(1+\gamma\right)\phi_{0}^{2}+\left(1+\gamma\right)\left(2+\gamma\right)\phi_{0}^{2+2\gamma},
\end{eqnarray}
and have used the fact that $\phi_{0}(r)$ is real. We note that,
the Bogoliubov equations in the above can be straightforwardly derived
by using the ansatz
\begin{equation}
\phi(\mathbf{r},\tau)=\phi_{0}(r)+\sum_{j}\left[u_{j}\left(\mathbf{r}\right)e^{-i\omega_{j}\tau}+v_{j}^{*}\left(\mathbf{r}\right)e^{+i\omega_{j}\tau}\right]
\end{equation}
to expand the time-dependent extended GPE Eq. (\ref{eq:dimensionlessEGPE})
in the first order in $u_{j}(\mathbf{r})$ and $v_{j}(\mathbf{r})$.
This linearization is a standard procedure to study the collective
density oscillations. It is also useful to note that, the operator
$\mathcal{\hat{M}}$ is related to the local compressibility of the
droplet, i.e., $n(\partial^{2}\epsilon/\partial n^{2})$ with $n=\phi_{0}^{2}$,
and the zero-frequency solution of the Bogoliubov equations is precisely
the condensate wave-function $\phi_{0}$, i.e., $u(\mathbf{r})=-v(\mathbf{r})=\phi_{0}(r)$.

The numerical workload of solving the stationary GPE and Bogoliubov
equations can be greatly reduced by exploiting the spherical symmetry
of the droplet. We first consider the solutions of the Schrödinger
equation 
\begin{equation}
\mathcal{\hat{L}}\psi_{l\alpha}(\mathbf{r})=\varepsilon_{l\alpha}\psi_{l\alpha}(\mathbf{r}),\label{eq:SchrodingerEQ}
\end{equation}
for a given good angular momentum quantum number $l$. As we shall
see in the next paragraph, this equation includes the non-linear stationary
GPE as a specific case, when we take the lowest energy state in the
$l=0$ sector as the condensate wave-function and self-consistently
solve the equation in an iterative way. In general, the wave-function
of Eq. (\ref{eq:SchrodingerEQ}) can be written as 
\begin{equation}
\psi_{l\alpha}\left(\mathbf{r}\right)=\frac{\Psi_{l\alpha}\left(r\right)}{r}Y_{lm}(\theta,\varphi)\label{eq:radialWF}
\end{equation}
with the spherical harmonics $Y_{lm}$, and the Schrödinger equation
changes into the form,
\begin{equation}
\left[-\frac{1}{2}\frac{d^{2}}{dr^{2}}+V_{\textrm{eff}}^{(l)}\left(r\right)\right]\Psi_{l\alpha}\left(r\right)=\varepsilon_{l\alpha}\Psi_{l\alpha}\left(r\right),\label{eq:radialSchrodingerEQ}
\end{equation}
where the effective potential
\begin{equation}
V_{\textrm{eff}}^{(l)}=\frac{\omega_{0}^{2}r^{2}}{2}+\frac{l\left(l+1\right)}{2r^{2}}-2\left(1+\gamma\right)\phi_{0}^{2}+\left(2+\gamma\right)\phi_{0}^{2+2\gamma}\label{eq:Veff}
\end{equation}
includes the original harmonic trapping potential, the centrifugal
potential, the attractive mean-field potential and the repulsive potential
from quantum fluctuations. This radial Schrödinger equation can be
conveniently solved by uniformly discretizing the $r$-axis with a
mesh length $\delta$ and by approximating $\Psi(r)$ by $\Psi_{p}$
for $p\delta<r\leq(p+1)\delta$ \citep{Pu1998}. Here, for clarity
we have dropped the indices $l$ and $\alpha$ in $\Psi(r)$. The
boundary conditions $\Psi_{p=0}=0$ and $\Psi_{p=M}=0$ can be applied
for a sufficiently large cut-off integer $M\gg1$. Taking the simplest
finite-difference approximation for the second derivative, i.e., $d^{2}\Psi(r)/dr^{2}=(\Psi_{p+1}-2\Psi_{p}+\Psi_{p-1})/\delta^{2}$,
we can represent the Hamiltonian in the radial Schrödinger equation
in a tridiagonal matrix of the form,
\begin{equation}
\left[\begin{array}{ccccc}
v_{\textrm{eff}}^{(l)}\left(1\right) & -\frac{1}{2}\delta^{-2}\\
-\frac{1}{2}\delta^{-2} & v_{\textrm{eff}}^{(l)}\left(2\right) & \ddots\\
 & \ddots & \ddots & \ddots\\
 &  & \ddots & v_{\textrm{eff}}^{(l)}\left(M-1\right) & -\frac{1}{2}\delta^{-2}\\
 &  &  & -\frac{1}{2}\delta^{-2} & v_{\textrm{eff}}^{(l)}\left(M\right)
\end{array}\right],
\end{equation}
where $v_{\textrm{eff}}^{(l)}(p)\equiv\delta^{-2}+V_{\textrm{eff}}^{(l)}(r=p\delta)$.
We choose a sufficiently small mesh size $\delta$, so that the relative
numerical error due to the finite-difference approximation for the
second derivative is about $10^{-4}$. By diagonalizing this real
symmetric matrix using the standard Eigenvalue Analysis Package, we
obtain $\varepsilon_{l\alpha}$ and the normalized radial wave-function
$\Psi_{l\alpha}(r)$.

We use such a procedure to solve the stationary extended GPE Eq. (\ref{eq:stationaryEGPE}),
for which $\phi_{0}(r)=\Psi_{l=0}(r)/(\sqrt{4\pi}r)$ is the lowest
eigenstate of the radial Schrödinger equation Eq. (\ref{eq:radialSchrodingerEQ})
with the corresponding eigenvalue being the chemical potential $\mu$.
As the condensate wave-function also appears in the effective potential
Eq. (\ref{eq:Veff}), iteration is necessary. To overcome the numerical
instability due to nonlinearity, we follow the strategy by Pu and
Bigelow in their seminal work on two-species Bose condensates \citep{Pu1998}
and introduce a controlling positive number $0<\vartheta<1$ to generate
a new wave-function for the next iteration. In detail, by using the
Gaussian variational wave-function to seed the start-up trial wave-function
$\Psi_{l=0}^{(0)}(r)$, for the $i$-th iteration we find the new
trial wave-function $\Psi_{l=0}^{(i)}(r)=\vartheta\Psi_{l=0}^{(i-1)}(r)+(1-\vartheta)\Psi_{l=0}(r)$,
where $\Psi_{l=0}(r)$ is the lowest-energy solution of Eq. (\ref{eq:radialSchrodingerEQ})
with $\phi_{0}=\sqrt{4\pi}r\Psi_{l=0}^{(i-1)}(r)$. The choice of
the controlling parameter $\vartheta$ depends on the nonlinearity
of the system: for large reduced number of particles $N\sim10^{4}$,
we find that it is necessary to set $\vartheta>0.95$, which requires
several hundred iterations for reaching the convergence (i.e., the
relative difference in the wave-functions before and after the final
iteration is less than $10^{-6}$).

To solve the Bogoliubov equations, we follow the work by Hutchinson,
Zaremba and Griffin \citep{Hutchinson1997}, and introduce the auxiliary
functions $\psi_{j\pm}(\mathbf{r})\equiv u_{j}(\mathbf{r})\pm v_{j}(\mathbf{r})$,
which satisfy the equations ($\mathcal{\hat{H}}_{0}\equiv\mathcal{\hat{L}}-\mu$),
\begin{eqnarray}
\mathcal{\hat{H}}_{0}\left[\mathcal{\hat{H}}_{0}+2\mathcal{\hat{M}}\right]\psi_{j+}\left(\mathbf{r}\right) & = & \omega_{j}^{2}\psi_{j+}\left(\mathbf{r}\right),\label{eq:FaiPlus}\\
\left[\mathcal{\hat{H}}_{0}+2\mathcal{\hat{M}}\right]\mathcal{\hat{H}}_{0}\psi_{j-}\left(\mathbf{r}\right) & = & \omega_{j}^{2}\psi_{j-}\left(\mathbf{r}\right),\label{eq:FaiMinus}
\end{eqnarray}
respectively. The two auxiliary functions are related to each other
by 
\begin{equation}
\mathcal{\hat{H}}_{0}\psi_{j-}\left(\mathbf{r}\right)=\omega_{j}\psi_{j+}\left(\mathbf{r}\right).\label{eq:FaiPMRelation}
\end{equation}
For a given angular momentum $l$, we then expand $\psi_{j-}(\mathbf{r})=\sum_{\alpha}c_{\alpha}^{(n)}\psi_{l\alpha}(\mathbf{r})$
in terms of the normalized eigenfunction basis $\psi_{l\alpha}(\mathbf{r})$
already solved in Eq. (\ref{eq:radialSchrodingerEQ}) and substitute
the expansion into Eq. (\ref{eq:FaiMinus}). It becomes clear that
the level index $j$ is represented by $j=(ln)$, where $n$ is the
radial quantum number. By recalling that $\mathcal{\hat{H}}_{0}\psi_{l\alpha}(\mathbf{r})=\tilde{\varepsilon}_{l\alpha}\psi_{l\alpha}(\mathbf{r})$
with 
\begin{equation}
\tilde{\varepsilon}_{l\alpha}=\varepsilon_{l\alpha}-\mu\geq0,\label{eq:energymu}
\end{equation}
we obtain the secular equation,
\begin{equation}
\sum_{\beta}\left[\tilde{\varepsilon}_{l\alpha}^{2}\delta_{\alpha\beta}+\sqrt{\tilde{\varepsilon}_{l\alpha}\tilde{\varepsilon}_{l\beta}}M_{\alpha\beta}\right]\sqrt{\tilde{\varepsilon}_{l\beta}}c_{\beta}^{(n)}=\omega_{ln}^{2}\sqrt{\tilde{\varepsilon}_{l\alpha}}c_{\alpha}^{(n)},
\end{equation}
where the matrix element $M_{\alpha\beta}\equiv2\int d\mathbf{r}\psi_{l\alpha}^{*}(\mathbf{r})\mathcal{\hat{M}}\psi_{l\beta}(\mathbf{r})$.
Therefore, once again we need to diagonalize a real symmetric matrix
to determine the eigenvalues $\omega_{ln}$ and the corresponding
coefficients $c_{\alpha}^{(n)}$. The latter is subject to the orthonormality
requirement of the $u_{j}$ and $v_{j}$ functions, i.e., $\int d\mathbf{r}[\left|u_{j}(\mathbf{r})\right|^{2}-\left|v_{j}(\mathbf{r})\right|^{2}]=1$,
which gives rise to the normalization condition, 
\begin{equation}
\sum_{\alpha}\tilde{\varepsilon}_{l\alpha}c_{\alpha}^{(n)}c_{\alpha}^{(n')}=\omega_{ln}\delta_{nn'}.
\end{equation}
Using the solution $\psi_{j-}(\mathbf{r})$, we now apply the relation
(\ref{eq:FaiPMRelation}) to obtain the auxiliary function $\psi_{j+}(\mathbf{r})$.
By combining these two functions, we find that,
\begin{equation}
\left\{ \begin{array}{c}
u_{ln}(\mathbf{r})\\
v_{ln}(\mathbf{r})
\end{array}\right\} =\frac{1}{2}\sum_{\alpha}\left[\frac{\tilde{\varepsilon}_{l\alpha}}{\omega_{ln}}\pm1\right]c_{\alpha}^{(n)}\phi_{l\alpha}\left(\mathbf{r}\right),
\end{equation}
and similar expressions for the radial functions $u_{ln}(r)$ and
$v_{ln}(r)$, following the correspondence between $\psi_{l\alpha}(\mathbf{r})$
and $\Psi_{l\alpha}(r)$ (see Eq. (\ref{eq:radialWF})).

In comparison with the solution of the Bogoliubov equations for a
weakly interacting Bose gas \citep{Hutchinson1997}, numerical calculations
for the droplet state are much more involved, due to the existence
of very low energy excitations and the large degeneracies in energy
close to the particle-emission threshold. The problem is particularly
severe for the dipole mode of a self-bound droplet. The \emph{zero}
frequency of the dipole mode without external harmonic traps can never
be exactly reproduced in our numerical calculations, since we have
to choose a cut-off integer $M$ with hard-wall boundary condition
to make calculation feasible. In practice, for self-bound droplet
we set the mesh size $\delta<0.1$ and choose $M\sim2000$. The largest
length in our calculations is therefore $r_{\textrm{max}}=M\delta\sim200$,
at least 20 times larger than the typical size of the droplet considered
in this work. The dipole mode frequency is typically about $10^{-3}$,
which provides an upper-bound estimate for the accuracy of the frequency
of other modes. In the presence of a harmonic trapping potential $\omega_{0}\geq0.01$,
the hard-wall boundary condition at $r_{\textrm{max}}$ is automatically
realized as the trapping potential $(\omega_{0}r_{\textrm{max}})^{2}/2\gg1$.
The relative deviation of the dipole mode frequency from $\omega_{0}$
is small and less than $1\%$. On the other hand, to overcome the
problem due to the large degeneracies close to the particle-emission
threshold, we use up to 500 eigenfunctions $\psi_{l\alpha}(\mathbf{r})$
as the expansion basis in solving the Bogoliubov equations. This number
is about 10 times larger than what is used for a weakly interacting
Bose condensate \citep{Hutchinson1997}.

\begin{figure}[t]
\begin{centering}
\includegraphics[width=0.48\textwidth]{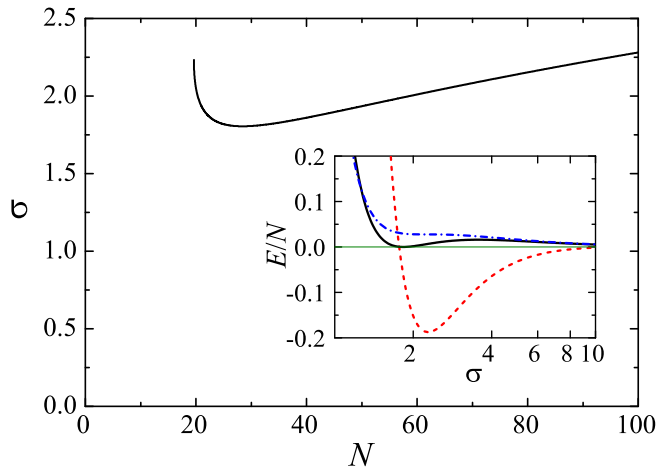}
\par\end{centering}
\caption{\label{fig1_GaussianWidth} The width of the Gaussian ansatz $\sigma$
as a function of the reduced number of particles $N$ at the $\gamma=1/2$
case. The inset shows the energy per particle predicted by the Gaussian
ansatz at the critical number $N_{c}$ (blue dot-dashed line), at
the threshold number for metastable states $N_{m}$ (black line),
and at the stable configuration $N=100$ (red dashed line).}
\end{figure}

\section{Self-bound quantum droplets in free space}

In this section, we consider a self-bound quantum droplet in free
space with $\omega_{0}=0$. 

\subsection{Critical particle numbers and phase diagram}

Let us start from the Gaussian variational approach, which provides
a qualitative description of the droplet state. As can be seen from
Eq. (\ref{eq:GaussianEtot}) \citep{Petrov2015}, the radius of the
droplet in free space $R\sim\sigma$ is set by competition among the
kinetic energy per particle $\epsilon_{\textrm{kin}}/N\propto1/\sigma^{2}$,
the mean-field energy per particle $\epsilon_{\textrm{MF}}/N\propto-N/\sigma^{3}$,
and the beyond-mean-field energy per particle $\epsilon_{\textrm{BMF}}/N\propto(N/\sigma^{3})^{1+\gamma}$.
The balance between the latter two energies gives rise to a finite
Gaussian width $\sigma\sim N^{^{1/3}}$. However, for small number
of particles the kinetic energy term quickly becomes dominant and
makes the droplet state unstable. Indeed, as shown in the inset of
Fig. \ref{fig1_GaussianWidth} for the $\gamma=1/2$ case, with decreasing
particle number the well-defined global minimum in the total energy
curve at large particle number (see, i.e., the red dashed line at
$N=100$) turns into a local minimum at a threshold $N_{m}$ ($\sim24$,
the black line), and the local minimum eventually disappears at a
slightly smaller critical number $N_{c}$ ($\sim19$, the blue dot-dashed
line). For the particle number at the interval $N\in[N_{c},N_{m}]$,
the system is metastable, as the particle may escape to free space
(with zero energy) through a finite energy barrier. This metastable
regime is typically correlated with an abnormal rapid increase in
the Gaussian width $\sigma$ when the particle number decreases, as
illustrated in Fig. \ref{fig1_GaussianWidth}.

\begin{figure}[t]
\begin{centering}
\includegraphics[width=0.48\textwidth]{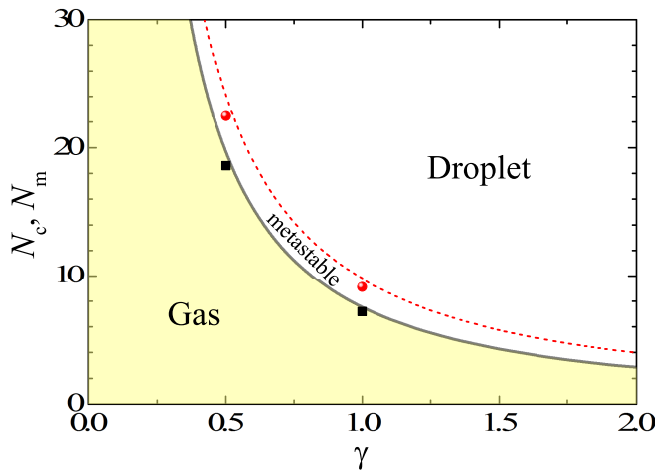}
\par\end{centering}
\caption{\label{fig2_NcNm} $N_{c}$ and $N_{m}$ as a function of the parameter
$\gamma$ predicted by the Gaussian ansatz, which give rise to a phase
diagram of the system at a given $\gamma$. The symbols show the results
from the extended GPE at $\gamma=1/2$ and $\gamma=1$.}
\end{figure}

The values of the critical and threshold particle numbers $N_{c}$
and $N_{m}$ could be calculated analytically with the Gaussian ansatz.
At $N=N_{c}$, as the local minimum starts to appear, the second derivative
of the total energy per particle with respect to $\sigma$ is zero,
$\partial^{2}(\epsilon_{\textrm{tot}}/N)/\partial\sigma^{2}=0$. Together
with the condition Eq. (\ref{eq:GaussianWidth}) at $\omega_{0}=0$,
we obtain,
\begin{equation}
N_{c}=\frac{2^{3\left(1+\gamma\right)/\left(4\gamma\right)}\left(1+3\gamma\right)^{\left(1+3\gamma\right)/\left(2\gamma\right)}\pi^{3/2}}{3^{3/2}\left(2+\gamma\right)^{3/\left(4\gamma\right)}\left[\gamma\left(1+\gamma\right)\right]^{3/2}}.
\end{equation}
On the other hand, at $N=N_{m}$, the disappearance of the local minimum
means a zero total energy $\epsilon_{\textrm{tot}}=0$ and we find
that,
\begin{equation}
\frac{N_{m}}{N_{c}}=\left(\frac{3}{2}\right)^{3/2}\left(1+\gamma\right)^{1/\left(2\gamma\right)}.
\end{equation}
For $\gamma$ increases from $0$ to $\infty$, this ratio increases
monotonically from $e^{-1/2}(3/2)^{3/2}\simeq1.1143$ to $(3/2)^{3/2}\simeq1.8371$.
Moreover, at $\gamma=1/2$ and $\gamma=1$, it takes values $\sqrt{3/2}\simeq1.2247$
and $3\sqrt{3}/4\simeq1.2990$, respectively.

In Fig. 2, we show $N_{c}$ and $N_{m}$ as a function of the parameter
$\gamma$, which provides a useful phase diagram for the system with
a finite particle number. The critical particle numbers obtained by
using the Gaussian ansatz (lines) and by solving the stationary extended
GPE (symbols) are remarkably close. At $\gamma=1/2$ ($\gamma=1$),
we find that the approximate Gaussian predictions $N_{c}\simeq19.62$
and $N_{m}\simeq24.03$ ($N_{c}\simeq7.52$ and $N_{m}\simeq9.77$)
are just a few percent larger than the exact numerical results from
the extended GPE $N_{c}\simeq18.65$ and $N_{m}\simeq22.55$ \citep{Petrov2015}
($N_{c}\simeq7.23$ and $N_{m}\simeq9.18$). The critical particle
numbers seem to increase significantly with decreasing $\gamma$.
This is deceptive, since the unit of the particle number $n_{0}\xi^{3}$
also depends on the parameter $\gamma$. Actually, we find that the
actual critical number of particles decreases with decreasing $\gamma$
, if we use the DMC equation of state \citep{Cikojevic2019,Cikojevic2020arXiv}.

\begin{figure}[t]
\begin{centering}
\includegraphics[width=0.48\textwidth]{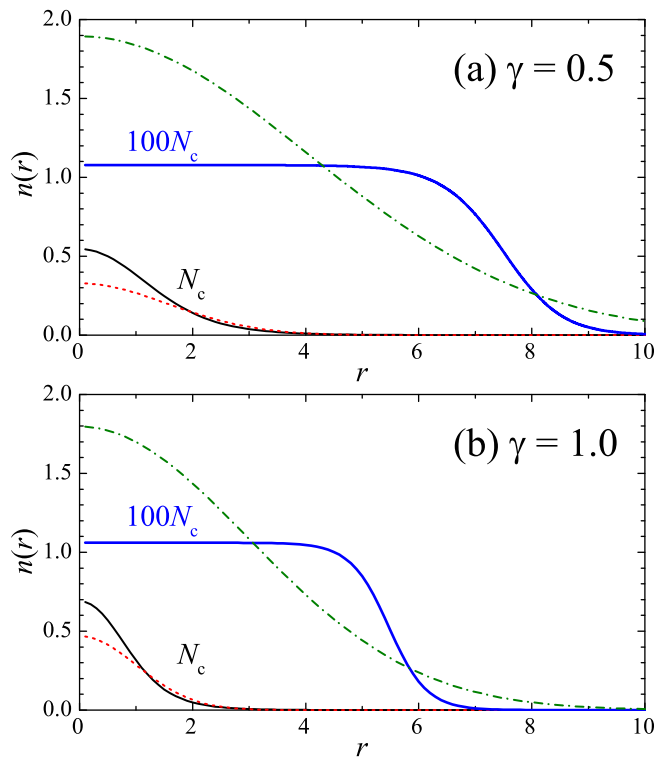}
\par\end{centering}
\caption{\label{fig3_GaussianDensity} The density profiles of a self-bound
droplet at $N=N_{c}$ and $N=100N_{c}$, with the parameter $\gamma=1/2$
(a, upper panel) and $\gamma=1$ (b, lower panel). The critical number
of particles is obtained by using the variational ansatz. The dashed
and dot-dashed lines show the predictions from the Gaussian ansatz
and the solid lines show the results of the extended GPE. The density
is measured in units of the equilibrium density $n_{0}$.}
\end{figure}

\begin{figure}[t]
\begin{centering}
\includegraphics[width=0.48\textwidth]{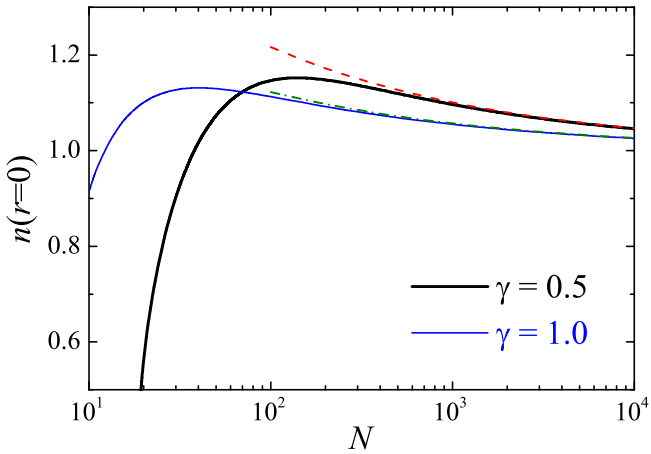}
\par\end{centering}
\caption{\label{fig4_n0} Central density of a self-bound droplet predicted
by the extended GPE at the parameter $\gamma=1/2$ (blue thin line)
and $\gamma=1$ (black thick line). The dashed and dot-dashed line
show the analytic results anticipated at large number of particles,
see Eq. (\ref{eq:n0}). The central density is measured in units of
the equilibrium density $n_{0}$.}
\end{figure}

\subsection{Density profile of the droplet}

The predictive power of the Gaussian variational approach becomes
worse if we consider the density distribution of a droplet. In Fig.
\ref{fig3_GaussianDensity}, we report the density profiles of a self-bound
droplet at the critical number of particles $N_{c}$ and $100N_{c}$,
calculated by using either the approximate Gaussian ansatz (dashed
or dot-dashed lines) or by solving the extend GPE (solid lines). While
there is a reasonable agreement between the predictions from the two
approaches at the small critical number $N_{c}$, the flat-top structure
of the droplet state at the relatively large particle number (i.e.,
$100N_{c}$) is completely missed by the Gaussian variational approach.

It is readily seen that the density of the flat-top part is larger
than unity, which we anticipate in the thermodynamic limit (i.e.,
$n_{r=0}=[\phi^{(\infty)}]^{2}=1$ in the re-scaled units). This deviation
is highlighted in Fig. \ref{fig4_n0}, where we show the central density
as a function of the number of particles $N$ at $\gamma=1/2$ and
$\gamma=1$. As $N$ increases, the central density initially rises
up rapidly, reaches a maximum at about $5N_{c}$ and finally saturates
very slowly towards the unity equilibrium density. The slow saturation
could be understood from the Laplace's formula for the surface pressure
(i.e., the pressure difference between the droplet and the surrounding
vacuum) \citep{FluidMechanicsBook1987},
\begin{equation}
P=\frac{2\sigma_{s}}{R},\label{eq:LaplaceFormula}
\end{equation}
where $\sigma_{s}=\lim_{\mathcal{S}\rightarrow\infty}[\epsilon_{\textrm{tot}}-\mu^{(\infty)}N]/\mathcal{S}$
is the surface tension, and $R\simeq[3N/(4\pi)]^{1/3}$ and $\mathcal{S}\equiv4\pi R^{2}$
are the radius and the surface area of the droplet, respectively.
In other words, due to the surface tension for a finite-size droplet,
the bulk pressure becomes nonzero. By recalling that in the re-scaled
units, the bulk pressure is given by 
\begin{equation}
P=\left[n\frac{\partial\epsilon}{\partial n}-\epsilon\right]_{n_{r=0}}=-\left(1+\gamma\right)n_{r=0}^{2}\left[1-n_{r=0}^{\gamma}\right],
\end{equation}
we find that,
\begin{equation}
n_{r=0}\simeq1+\frac{2\sigma_{s}}{\gamma\left(1+\gamma\right)}\left(\frac{4\pi}{3}\right)^{1/3}N^{-1/3}.\label{eq:n0}
\end{equation}
Following Stringari and Treiner \citep{Stringari1987}, the surface
tension can be written as ($x=r-R$),
\begin{align}
\sigma_{s} & =\intop_{-R}^{\infty}dx\left[\frac{1}{2}\left(\frac{d\phi_{0}}{dx}\right)^{2}+\epsilon\left(\phi_{0},\phi_{0}\right)-\mu^{(\infty)}\phi_{0}^{2}\right],\\
 & =\frac{1}{\sqrt{2}}\intop_{0}^{1}dn\left[-\left(1+\gamma\right)n+n^{\left(1+\gamma\right)}+\gamma\right]^{1/2},
\end{align}
where in the second line we have used the equation of motion 
\begin{equation}
-\frac{1}{2}\left(\frac{d\phi_{0}}{dx}\right)^{2}-\left(1+\gamma\right)\phi_{0}^{4}+\phi_{0}^{4+2\gamma}=\mu^{(\infty)}\phi_{0}^{2}
\end{equation}
valid in the thermodynamic limit to convert $dx$ to $dn$, where
$n\equiv\phi_{0}^{2}(x)$. It is then straightforward to perform the
integration, and we find $\sigma_{s}(\gamma=1/2)=3(1+\sqrt{3})/35\simeq0.234176$
\citep{Petrov2015} and $\sigma_{s}(\gamma=1)=2^{-3/2}\simeq0.353553$.
The surface tension increases with increasing $\gamma$. For sufficiently
large $\gamma$, it takes the form,
\begin{equation}
\sigma_{s}\left(\gamma\rightarrow\infty\right)\simeq\frac{\sqrt{2\gamma}}{3}.
\end{equation}
In Fig. \ref{fig4_n0}, we show the asymptotic behavior Eq. (\ref{eq:n0})
of the central density by using dashed and dot-dashed lines, for $\gamma=1/2$
and $\gamma=1$, respectively. At large particle number $N>10^{3}$,
the asymptotic relation works extremely well.

\begin{figure}[t]
\begin{centering}
\includegraphics[width=0.48\textwidth]{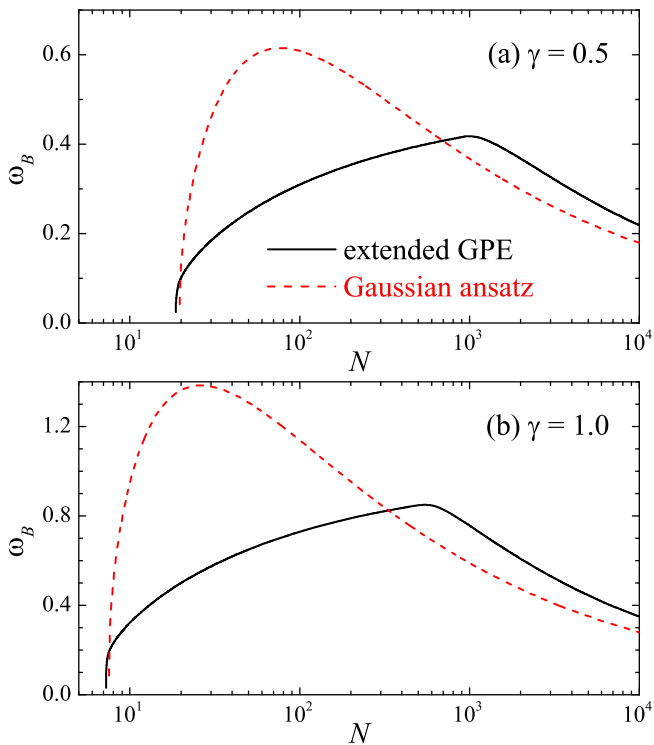}
\par\end{centering}
\caption{\label{fig5_GaussianWB} Breathing mode frequency $\omega_{B}$ of
a self-bound droplet predicted by the Gaussian ansatz (red dashed
lines) and the extended GPE (black solid lines) at the parameters
$\gamma=1/2$ (a, upper panel) and $\gamma=1$ (b, low panel).}
\end{figure}

\subsection{Collective excitations of the droplet}

We now turn to consider the collective excitations of a self-bound
quantum droplet, starting from the lowest monopole mode, the breathing
mode. In Fig. \ref{fig5_GaussianWB}, we show breathing mode frequencies
as a function of the particle number at $\gamma=1/2$ (a) and $\gamma=1$
(b), predicted by the Gaussian variational approach (dashed line)
and by the Bogoliubov equations (solid lines). In sharp contrast to
the one-dimensional case, where there is a good agreement between
the results from the two methods \citep{Astrakharchik2018}, here
we find that the Gaussian variational approach strongly over-estimates
the breathing mode frequency for small number of particles and incorrectly
predicts a large peak at $N\sim4N_{c}$. Only at very large particle
number, i.e., $N>50N_{c}$, the variational ansatz begins to provide
qualitatively correct mode frequency $\omega_{B}\propto\sqrt{N/\sigma^{5}}\propto1/R$.
This decrease in the breathing mode frequency at large number of particles
is anticipated. As we shall discuss later, the breathing mode is the
lowest compressional sound mode, whose frequency is given by $\omega_{B}\simeq ck_{B}$,
where $c$ is the bulk sound velocity and $k_{B}$ is the characteristic
wave-vector of the breathing mode. As $k_{B}$ is inversely proportional
to the radius of the droplet, i.e., $k_{B}\propto1/R$, the breathing
mode frequency $\omega_{B}\propto1/R$ then has to decrease, when
the radius of the droplet becomes larger. This leads to a peak in
the breathing mode frequency as a function of the number of particles,
as shown by the red dashed lines in Fig. \ref{fig5_GaussianWB}. 

\begin{figure}[t]
\begin{centering}
\includegraphics[width=0.48\textwidth]{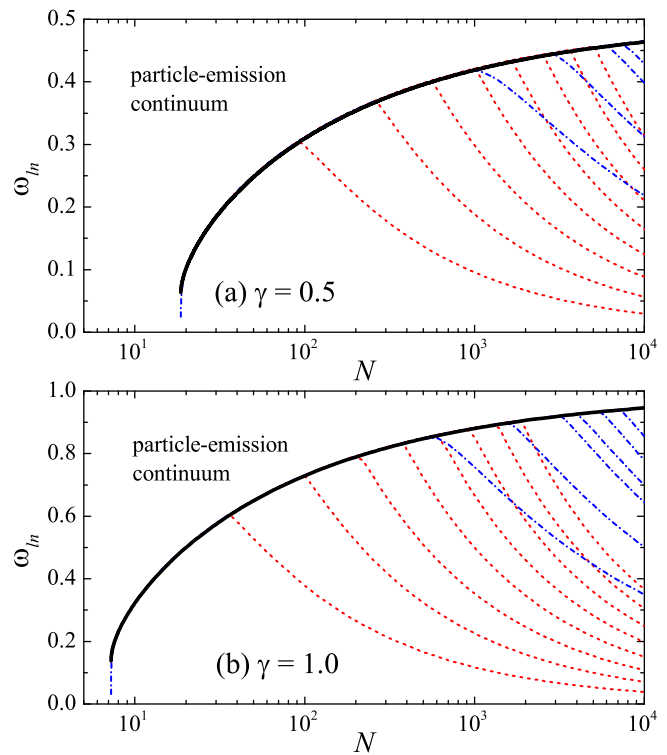}
\par\end{centering}
\caption{\label{fig6_CollectiveModes} Excitation frequencies $\omega_{ln}$
($l\protect\leq9$ and $n\protect\leq2$) of a self-bound droplet,
as a function of the reduced particle number $N$ at the parameter
$\gamma=1/2$ (a, upper panel) and $\gamma=1$ (b, lower panel). The
red dashed lines show the surface modes $\omega_{l\protect\geq2,n=0}$
and the blue dot-dashed lines shows the other bulk modes. The black
thick lines plot the particle-emission continuum $-\mu$, above which
the excitations become unbound and acquire a free-particle dispersion
relation.}
\end{figure}

\begin{figure}[t]
\begin{centering}
\includegraphics[width=0.48\textwidth]{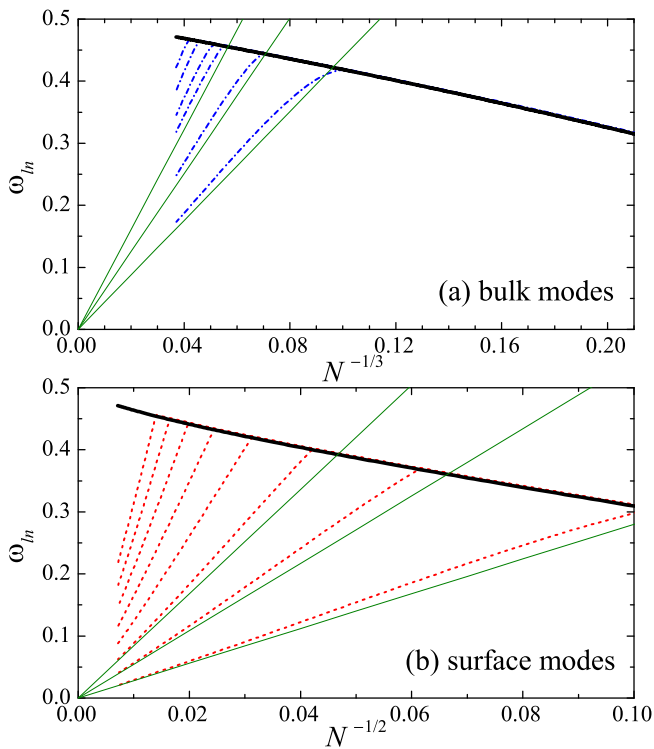}
\par\end{centering}
\caption{\label{fig7_CMScaling} Re-plot of the excitation frequencies $\omega_{ln}$
($l\protect\leq9$ and $n\protect\leq2$) of a self-bound droplet
at the parameter $\gamma=1/2$, as a function of \emph{$N^{-1/3}$}
(a, upper panel with bulk modes only) and $N^{-1/2}$ (b, lower panel
with surface modes only). The green straight lines show the analytic
results anticipated for a large droplet.}
\end{figure}

In Fig. \ref{fig6_CollectiveModes}, we present the whole spectrum
of collective excitations $\omega_{ln}$ ($l\leq9$ and $n\leq2$)
as a function of the particle number $N$ at the parameter $\gamma=1/2$
(a) and $\gamma=1$ (b). Basically, Fig. \ref{fig6_CollectiveModes}(a)
re-plots the mode frequencies found earlier by Petrov (see Fig. 1(b)
in Ref. \citep{Petrov2015}), but with a minor difference. That is,
we consider a larger regime for the number of particles, so the mode
frequencies $\omega_{ln}$ with nonzero radial quantum numbers $n\neq0$
start to show up. In the figure, those mode frequencies are plotted
by using blue dot-dashed lines, together with the breathing mode frequency
$\omega_{00}$. In contrast, the mode frequencies with $l\geq2$ and
$n=0$ are shown by using red dashed lines. The different illustration
of the modes comes from their different classification and characters.
The former is the so-called bulk mode, which is basically the sound
mode spreading throughout the whole droplet; while the latter is referred
to as the surface mode that only propagates at the edge of the droplet
and uses the surface tension as the restoring force for propagation
\citep{FluidMechanicsBook1987}. These two kinds of modes only survive
below the particle-emission threshold, i.e., $\omega_{ln}\leq-\mu$.
To understand this, let us recall Eq. (\ref{eq:energymu}) and note
that the quasi-particle energy $\omega_{ln}$ is measured with respect
to the chemical potential $\mu$, instead of the energy of the surrounding
vacuum (which is zero). In the frame of the vacuum, the actual energy
of the quasi-particle is then $\omega_{ln}+\mu>0$, if the excitation
energy $\omega_{ln}$ is above the particle-emission threshold. Therefore,
the quasi-particle will tunnel into the vacuum and acquires the free-particle
dispersion relation (i.e., the continuum). In other words, the negative
chemical potential of the droplet provides an effective confining
potential to quasi-particles within the droplet and this leads to
the discrete bulk and surface modes below the particle-emission threshold. 

Because of their different characters, the frequencies of the bulk
and surface modes have distinct dependences on the reduced number
of particles. For the bulk modes, if we approximate the droplet as
a ball with a sharp edge and radius $R$, we may write
\begin{equation}
\omega_{ln}^{(\textrm{bulk})}\simeq ck_{ln}=c\frac{z_{ln}}{R}\simeq cz_{ln}\left(\frac{4\pi}{3}\right)^{1/3}N^{-1/3},\label{eq:wbulk}
\end{equation}
where in the re-scaled units the speed of sound $c=\sqrt{(\partial P/\partial n)/m}$
is given by
\begin{equation}
c=\sqrt{\gamma\left(1+\gamma\right)}
\end{equation}
and $k_{ln}$ is the wave-vector of the mode satisfying the hard-wall
boundary condition $j_{l}(k_{ln}R)=0$, where $j_{l}(x)$ is the spherical
Bessel function of the first kind with zeros $z_{ln}$. For $l=0$,
the zeros are given by $z_{0n}=(n+1)\pi$. For nonzero angular momentum,
we have $z_{11}\simeq4.4934$, $z_{12}\simeq7.7253$, $\cdots$, $z_{21}\simeq5.7635$,
and so on. On the other hand, the dispersion relation of the surface
modes can be obtained by solving a Laplace's equation for the velocity
field with a boundary condition set by Eq. (\ref{eq:LaplaceFormula}),
i.e., $\omega_{l0}^{2}=l(l-1)(l+2)\sigma_{s}/(n_{r=0}mR^{3})$ \citep{FluidMechanicsBook1987},
which in the re-scale units takes the form,
\begin{equation}
\omega_{l0}^{(\textrm{surface})}\simeq\sqrt{\frac{4\pi l\left(l-1\right)\left(l+2\right)\sigma_{s}}{3}}N^{-1/2}.\label{eq:wsurface}
\end{equation}
It is clear that the bulk and surface mode frequencies scale like
$N^{-1/3}$ and $N^{-1/2}$, respectively. To highlight those different
scaling behaviors, in Fig. \ref{fig7_CMScaling} we re-plot the mode
frequencies as a function of $N^{-1/3}$ and $N^{-1/2}$ in (a) and
(b), respectively. The analytic predictions by Eq. (\ref{eq:wbulk})
and Eq. (\ref{eq:wsurface}) are also shown by thin green lines for
the lowest three modes. We find an excellent agreement between the
numerical and analytical results for sufficiently large number of
particles. The agreement for the surface modes is particularly satisfactory,
presumably due to their lower energy that favors the application of
the hydrodynamic equations.

\begin{figure}[t]
\begin{centering}
\includegraphics[width=0.48\textwidth]{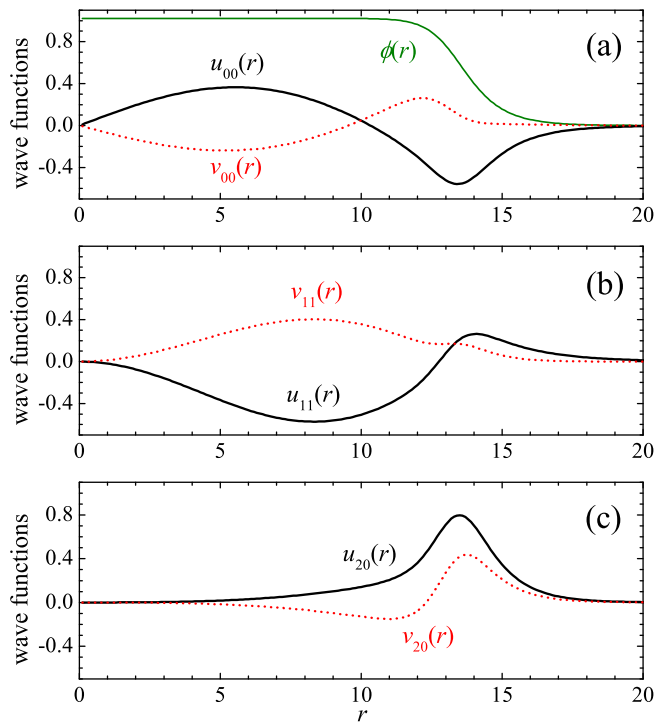}
\par\end{centering}
\caption{\label{fig8_uvLargeN} The quasi-particle wave-functions $u_{ln}(r)$
(black solid lines) and $v_{ln}(r)$ (red dashed lines) at a large
reduced particle number $N=10000$ and at the parameter $\gamma=1/2$.
We show the two lowest non-trivial bulk modes (a, $l=n=0$ and b,
$l=n=1$) and the lowest surface mode (c, $l=2$ and $n=0$). In (a),
we show also the condensate wave-function $\phi(r)$ of the free-bound
droplet. We note that, in (a) the wave-functions of the breathing
mode $u_{00}(r)$ and $v_{00}(r)$ have a node in the radial direction,
due to our definition of the radial quantum number $n$ \citep{NoteNumberNodes}.}
\end{figure}

\begin{figure}[t]
\begin{centering}
\includegraphics[width=0.48\textwidth]{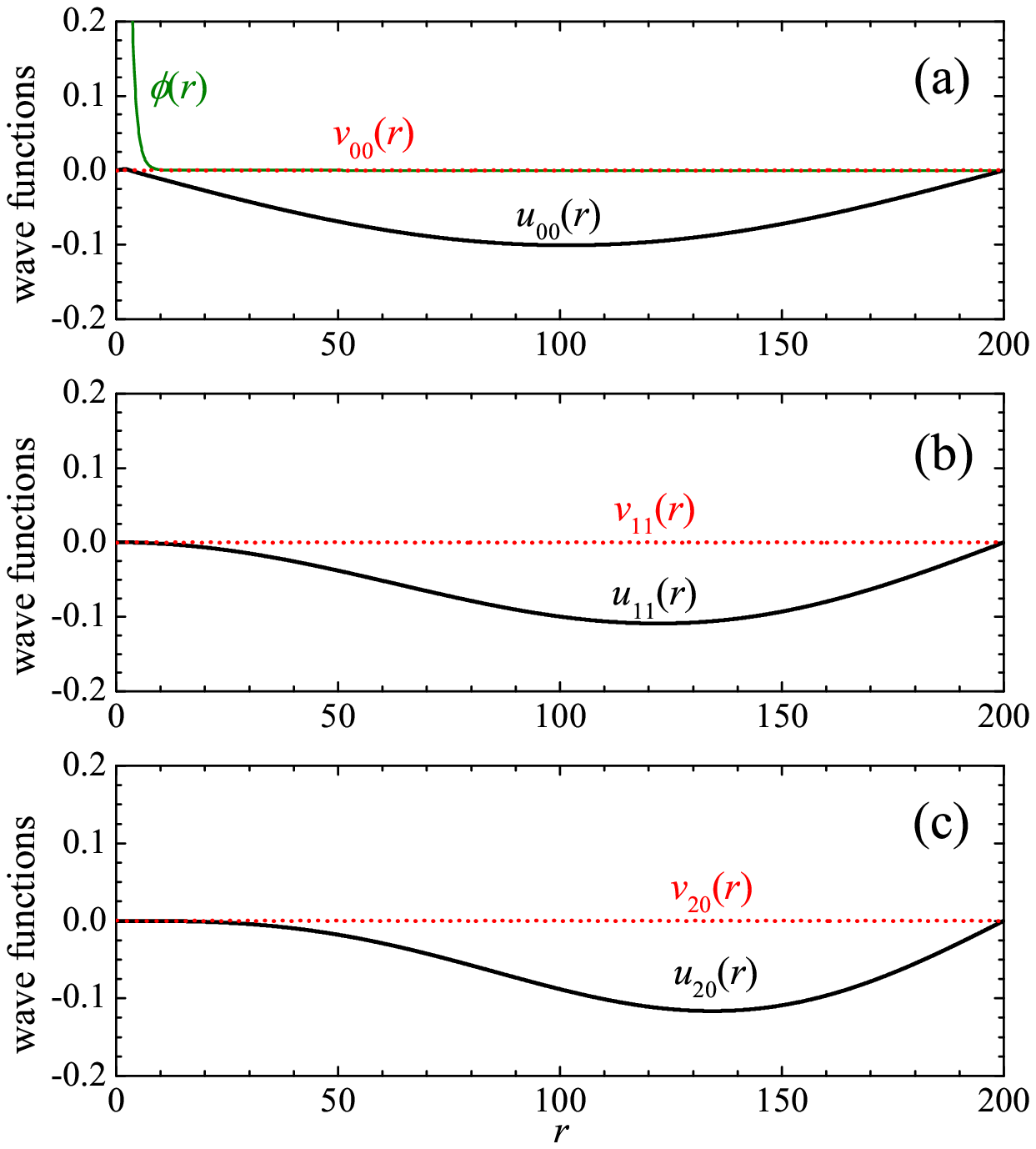}
\par\end{centering}
\caption{\label{fig9_uvSmallN} The quasi-particle wave-functions $u_{ln}(r)$
(black solid lines) and $v_{ln}(r)$ (red dashed lines) at a small
reduced particle number $N=60$ and at the parameter $\gamma=1/2$.
We show the two lowest non-trivial bulk modes (a, $l=n=0$ and b,
$l=n=1$) and the lowest surface mode (c, $l=2$ and $n=0$). In (a),
we show also the condensate wave-function $\phi(r)$ of the free-bound
droplet. We note that, the cut-off length in our numerical calculations
is $r_{\textrm{max}}\simeq200$.}
\end{figure}

The different characters of the bulk and surface modes might also
be understood from the quasi-particle wave-functions $u_{ln}(r)$
and $v_{ln}(r)$, which can be experimentally probed by measuring
the density fluctuation $\delta n(r)\sim[u_{ln}(r)+v_{ln}(r)]\phi_{0}(r)$.
In Fig. \ref{fig8_uvLargeN}, we show $u_{ln}(r)$ and $v_{ln}(r)$
at $\gamma=1/2$ and at a large number of particles $N=10000$, so
the bulk mode frequencies $\omega_{00}$ (a) and $\omega_{11}$ (b),
and the surface mode frequency $\omega_{20}$ (c) are all below the
particle-emission continuum. In comparison with the condensate wave-function
$\phi_{0}(r)$, i.e., the green line in Fig. \ref{fig8_uvLargeN}(a),
it is evident that the wave-functions of the bulk modes fluctuate
within the whole droplet, while the wave-functions of the surface
modes localize near the edge of the droplet only. Outside the droplet,
all the wave-functions decay exponentially. Thus, for the surface
modes, the excitations mainly perturb the density in the surface region,
as we anticipate.

Interestingly, for small number of particles, there is a threshold
$N_{\textrm{th}}$, below which all the excitation modes of a \emph{stable}
droplet in the interval $N_{m}<N<N_{\textrm{th}}$ lie above the particle-emission
threshold $-\mu$ (see Fig. \ref{fig6_CollectiveModes}). For $\gamma=1/2$
and $\gamma=1$, we find that $N_{\textrm{th}}\simeq94.2$ \citep{Petrov2015}
and $N_{\textrm{th}}\simeq36.5$, respectively. As pointed out by
Petrov \citep{Petrov2015}, in such an interval, the droplet fails
to create bound excitations and therefore cannot dissipate the energy
added to the system. In other words, upon excitations the droplet
needs to emit particles and evaporate automatically. This self-evaporation
phenomenon has recently been simulated by Ferioli and co-workers,
by preparing the droplet slightly out of equilibrium and consequently
monitoring the evolution of the droplet size \citep{Ferioli2020}
(see also the work \citep{Cikojevic2020arXiv}). The oscillation in
the size is found to decay quickly and the breathing mode frequency
extracted from the simulations decreases in time until it touches
the particle-emission threshold. 

In solving the Bogoliubov equations, the self-evaporation mechanism
can alternatively be understood from the quasi-particle wave-functions
$u_{ln}(r)$ and $v_{ln}(r)$, as shown in Fig. \ref{fig9_uvSmallN},
for the number of particles $N=60<N_{\textrm{th}}$ at the parameter
$\gamma=1/2$. We find that, for all the bulk and surface modes (considered
in the figure), the hole component of the wave-functions $v_{ln}(r)$
disappears, and the particle component $u_{ln}(r)$ becomes unbound,
in the sense that the shape of $u_{ln}(r)$ becomes completely irrelevant
to the droplet and is set by the maximum cut-off length $r_{\textrm{max}}\simeq200$
considered in the numerical calculations. This exactly implies the
emission of particles upon excitations.

For a general parameter $\gamma$, the threshold number $N_{\textrm{th}}$
might be analytically determined, by tracing the crossing point between
the lowest surface mode frequency $\omega_{20}$ and the particle-emission
threshold $-\mu$. By using Eq. (\ref{eq:wsurface}), we find $\omega_{20}=4\sqrt{2\pi\sigma_{s}/(3N)}$.
For the chemical potential, we note that, in the re-scaled units the
total energy is approximately given by $\epsilon_{\textrm{tot}}/N\simeq-\gamma+\epsilon_{s}N^{-1/3}$,
where $\epsilon_{s}=(36\pi)^{1/3}\sigma_{s}$ is the surface energy
\citep{Stringari1987}. Hence, we obtain $\mu=\partial\epsilon_{\textrm{tot}}/\partial N=-\gamma+(32\pi/3)^{1/3}\sigma_{s}N^{-1/3}$.
This $N^{-1/3}$-dependence of the chemical potential can also be
seen from the particle-emission threshold in Fig. \ref{fig7_CMScaling}(a).
By taking $\omega_{20}=-\mu$ at the threshold number of particles,
we find that
\begin{equation}
N_{\textrm{th}}\simeq\frac{32\pi\sigma_{s}}{3\left[\gamma-\left(\gamma\sigma_{s}\right)^{2/3}\right]^{2}}.
\end{equation}
This analytic expression predicts $N_{\textrm{th}}\simeq115.5$ and
$N_{\textrm{th}}\simeq47.4$ at $\gamma=1/2$ and $\gamma=1$, respectively,
which are about $20-30\%$ larger than the numerical results.

\begin{figure}[t]
\begin{centering}
\includegraphics[width=0.48\textwidth]{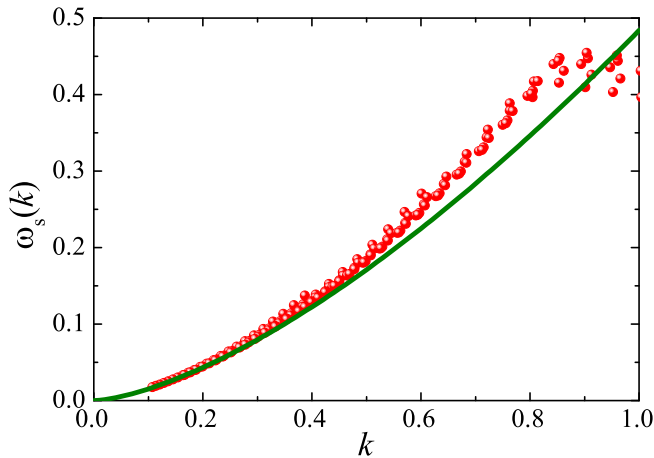}
\par\end{centering}
\caption{\label{fig10_wk15} Excitation frequencies of the surface modes $\omega_{l,n=0}$
from $l=2$ to $l=9$, at the parameter $\gamma=1/2$ for some selected
reduced particle numbers $N$ ranging from $500$ to $30000$, as
a function of the effective wave-vector $k=[l(l-1)(l+2)]^{1/3}/R$.
The green thick line shows the anticipated dispersion relation $\omega_{s}(k)=\sqrt{\sigma_{s}}k^{3/2}$,
where the dimensionless surface tension $\sigma_{s}=6(1+\sqrt{3})/35\simeq0.234176$
at the parameter $\gamma=1/2$.}
\end{figure}

The existence of the surface modes is a unique feature of quantum
droplets. By examining more closely Eq. (\ref{eq:wsurface}), it seems
useful to define an effective wave-vector \citep{Chin1995}
\begin{equation}
k=\frac{\left[l\left(l-1\right)\left(l+2\right)\right]^{1/3}}{R}
\end{equation}
 and re-cast the dispersion relation into the form, 
\begin{equation}
\omega_{s}\left(k\right)=\sqrt{\sigma_{s}}k^{3/2}.\label{eq:gk15}
\end{equation}
As the discreteness of the excitation spectrum becomes less important
for large droplets, we anticipate that such an exotic $k^{3/2}$ dispersion
relation should be valid at sufficiently large number of particles.
We have calculated the surface mode frequencies with $2\leq l\leq9$
for a large droplet with $N$ ranging from $500$ to $30000$, and
have taken the root-mean-square (rms) radius
\begin{equation}
R=\sqrt{\frac{5}{3}\left\langle r^{2}\right\rangle }\label{eq:Radius}
\end{equation}
to reduce the finite-size effect. The surface mode frequencies are
plotted as a function of the effective wave-vector $k$ in Fig. \ref{fig10_wk15}.
Indeed, we observe that the data points of the mode frequencies nicely
collapse onto the predicted dispersion relation Eq. (\ref{eq:gk15}),
when the number of particles becomes sufficiently large or the effective
wave-vector becomes sufficiently small. At larger effective wave-vector
(i.e., $k>0.8$), the curve of the data points turns out to abruptly
become flat. This is caused by either a small number of particles
or a large angular momentum $l$, at which the surface mode frequency
starts to merge with the particle-emission threshold $\left|\mu\right|$,
so the mode frequency $\omega_{l0}$ can no longer be described by
the ripplon dispersion Eq. (\ref{eq:wsurface}). Experimentally, the
frequency of the quadrupole surface mode $\omega_{20}$ could be readily
measured, together with the rms radius $R$ of the droplet, at different
large number of particles. As a result, the dispersion relation Eq.
(\ref{eq:gk15}) might be verified and the surface tension $\sigma_{s}$
is then experimentally determined.

\section{Quantum droplets in harmonic traps}

Let us now turn to consider the external harmonic trapping potential.
We will focus on the case with the parameter $\gamma=1/2$.

\subsection{$\omega_{0}$-dependence of the droplet profile and collective modes }

\begin{figure}[t]
\begin{centering}
\includegraphics[width=0.48\textwidth]{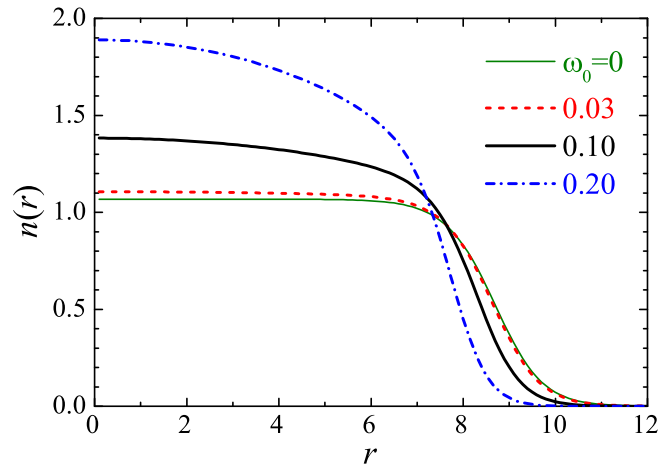}
\par\end{centering}
\caption{\label{fig11_DensityTrap} The density profiles of a droplet with
and without the harmonic trapping potential at the reduced particle
number $N=3000$ and at the parameter $\gamma=1/2$.}
\end{figure}

\begin{figure}[t]
\begin{centering}
\includegraphics[width=0.48\textwidth]{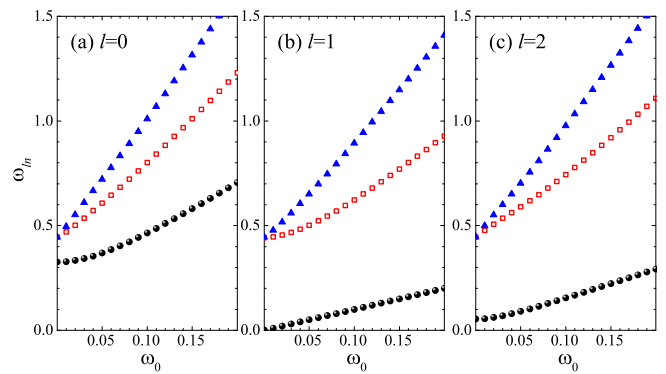}
\par\end{centering}
\caption{\label{fig12_w0dep} The dependence of excitation frequencies on the
external harmonic trapping potential: (a) $l=0$, (b) $l=1$ and (c)
$l=2$. In each panel, $n=0$, $n=1$ and $n=2$ from bottom to top.
We take the reduced number of particles $N=3000$ the parameter $\gamma=1/2$. }
\end{figure}

For a droplet with the size $R\simeq[3N/(4\pi)]^{1/3}$, qualitatively
we may anticipate the effect of the external trapping potential will
become important once the potential energy at the droplet edge $\omega_{0}^{2}R^{2}/2$
becomes comparable with the ``binding energy'' of the droplet $\left|\mu\right|$
in the absence of the trap, i.e., $\omega_{0}^{2}R^{2}/2\sim\left|\mu^{(\infty)}\right|=\gamma$,
which gives rise to a characteristic trapping frequency,
\begin{equation}
\omega_{0,c}\sim\sqrt{2\gamma}\left(\frac{4\pi}{3}\right)^{1/3}N^{-1/3}.
\end{equation}
In Fig. \ref{fig11_DensityTrap}, we report the density profiles of
the droplet with the number of particles $N=3000$ at different trapping
frequencies as indicated. In this case, the characteristic trapping
frequency $\omega_{0,c}\sim0.10$. Indeed, we find that the density
profiles at $\omega_{0}=0.03\ll\omega_{0,c}$ only differs sightly
from the self-bound droplet (at $\omega_{0}=0$). While at the trapping
frequency $\omega_{0}=0.20>\omega_{0,c}$, there is a significant
modulation to the density distribution due to the trapping potential.
The flat-top structure is lost and the central density deviates notably
from the equilibrium density in free space, i.e., $n_{r=0}=1$, as
we expect for a large self-bound droplet. As we shall discuss in detail
in the next section, the choice of a dimensionless trapping frequency
$\omega_{0}=0.03$ follows roughly the experiment conditions in Ref.
\citep{Semeghini2018}, where there is a residual trapping frequency
$\omega_{z}\sim2\pi\times12$ Hz along the axial direction. The negligible
trapping effect at $\omega_{0}=0.03$ shown in Fig. \ref{fig11_DensityTrap}
therefore strongly supports the claim that a self-bound droplet has
been observed in free space \citep{Semeghini2018}.

\begin{figure}[t]
\begin{centering}
\includegraphics[width=0.48\textwidth]{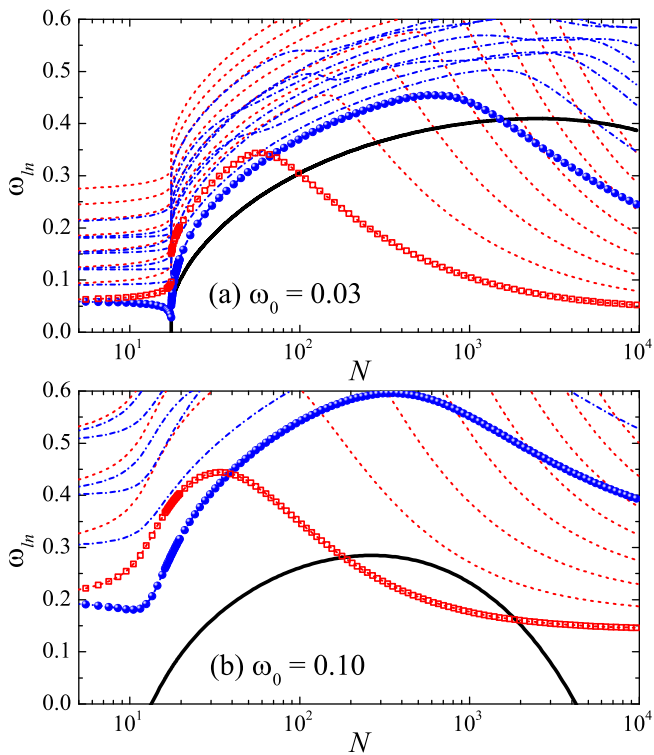}
\par\end{centering}
\caption{\label{fig13_CMTrap} Excitation frequencies $\omega_{ln}$ ($l\protect\leq9$
and $n\protect\leq2$) of an ultradilute droplet in harmonic traps,
as a function of the reduced particle number $N$ at the dimensionless
trapping frequency $\omega_{0}=0.03$ (a, upper panel) and $\omega_{0}=0.10$
(b, lower panel). The red dashed lines show the surface modes $\omega_{l\protect\geq2,n=0}$
and the blue dot-dashed lines show other bulk modes. The lowest surface
mode $\omega_{20}$ (i.e, quadruple mode) and the lowest bulk mode
(breathing monopole mode) are emphasized by the red open squares and
blue solid circles, respectively. The black thick lines show $-\mu$.
Here, we take the parameter $\gamma=1/2$.}
\end{figure}

In Fig. \ref{fig12_w0dep}, we show the frequencies $\omega_{ln}$
of the lowest three monopole (a), dipole (b) and quadrupole modes
(c) as a function of the trapping frequency $\omega_{0}$, for a droplet
with a large number of particles $N=3000$, at which in the self-bound
limit ($\omega_{0}\rightarrow0$) the discrete modes $\omega_{00}$
and $\omega_{20}$ already show up below the particle-emission threshold.
The lowest dipole mode is trivial and its mode frequency is always
the trapping frequency, $\omega_{10}=\omega_{0}$, owing to the well-known
Kohn theorem that under the harmonic trapping potential the center-of-mass
motion is an exact excited state of the system. The frequencies of
the breathing mode (i.e., the lowest monopole mode) and of the $l=2$
surface mode (i.e., the lowest quadruple mode) clearly show a \emph{super-linear}
dependence on $\omega_{0}$, indicating that those modes are not so
sensitive to the small trapping frequency satisfying $\omega_{0}<\omega_{0,c}$.
In contrast, for the higher excitation modes (i.e., those with $n=2$),
we typically find a linear dependence of the mode frequency on $\omega_{0}$.
This somehow suggests that those modes are mainly related to the external
trapping potential, instead of the intrinsic properties of the droplet.
In fact, in the self-bound limit those modes enter the particle-emission
continuum and become uncorrelated with the droplet.

\subsection{Excitation spectrum and phase diagram}

\begin{figure}[t]
\begin{centering}
\includegraphics[width=0.48\textwidth]{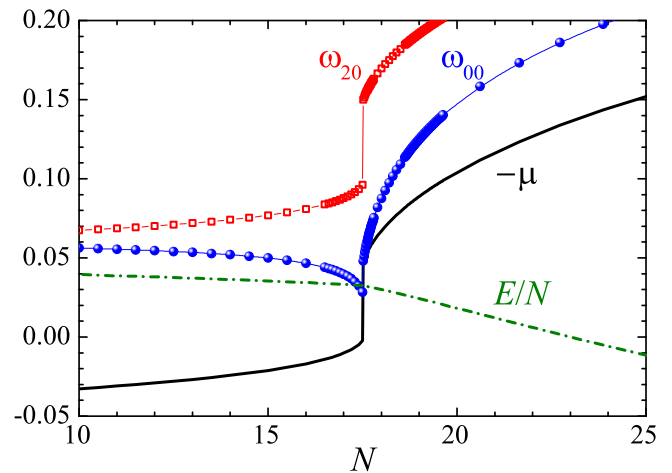}
\par\end{centering}
\caption{\label{fig14_CMTrapZoomIn} The enlarged view of the breathing and
quadruple excitation frequencies in Fig. \ref{fig13_CMTrap}(a), $\omega_{00}$
and $\omega_{20}$, near the critical number of particles. The black
thick line shows $-\mu$ and the green dot-dashed line corresponds
to the energy per particle $E/N$. There is a first-order quantum
phase transition from the droplet state to a gas-like state at $N\simeq17.5$.}
\end{figure}

In Fig. \ref{fig13_CMTrap}, we report the whole excitation spectrum
$\omega_{ln}$ as a function of the number of particles $N$, at $\omega_{0}=0.03$
(a) and $\omega_{0}=0.10$ (b). We show also the particle-emission
threshold $-\mu$ in black thick lines, although it becomes less well-defined
in the presence of an external harmonic trap. Following the earlier
convention, we have plotted the bulk and surface modes by using blue
dot-dashed lines and red dashed lines. Furthermore, the breathing
mode and the lowest $l=2$ surface mode are highlighted using blue
circles and red squares, respectively. Three features of the figure
are worth noting.

First, due to the existence of the external trapping potential, the
excitations become all bound and have discrete mode frequency. There
is no longer the particle-emission continuum. In particular, the system
now can have arbitrarily small number of particles, since the trapping
potential plays the role of container to confine particles in the
gas-like state. In the limit of vanishing number of particles, $N\rightarrow0$,
the system is basically a non-interacting gas, so the excitation spectrum
can be easily understood. For instance, the frequencies of the breathing
mode and the $l=2$ surface mode reach the non-interacting value $\omega_{B}=\omega_{Q}=2\omega_{0}$.
On the other hand, in the large particle number limit, $N\rightarrow\infty$,
the frequency of each mode tend to a finite value. This is particularly
clear for the lowest surface mode, whose frequency gradually approaches
$\omega_{Q}=\sqrt{2}\omega_{0}$, which is anticipated for interacting
quantum gases \citep{Dalfovo1999,Hu2004}.

\begin{figure}[t]
\begin{centering}
\includegraphics[width=0.48\textwidth]{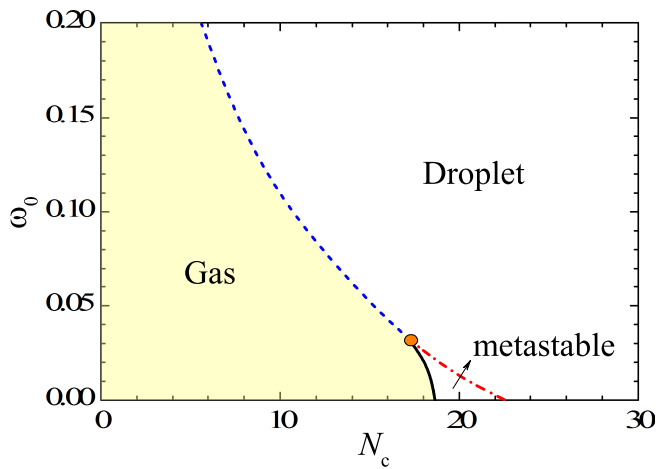}
\par\end{centering}
\caption{\label{fig15_PhaseDiagram} Phase diagram of a harmonically trapped
system near the gas-liquid transition at the parameter $\gamma=1/2$.
The solid line shows the first-order transition, while the dashed
or dot-dashed line indicates a smooth transition. The window for metastable
states shrinks with increasing trapping potential and disappears at
the dimensionless trapping frequency $\omega_{0}\simeq0.032$, as
indicated by the orange circle.}
\end{figure}

Second, at small trapping frequency (see Fig. \ref{fig13_CMTrap}(a)),
we typically find a jump in all the mode frequencies at a critical
number of particles. This is highlighted in Fig. \ref{fig14_CMTrapZoomIn},
where at $N_{c}\sim17.5$ we also find a discontinuity in the chemical
potential $-\mu$ and a kink in the energy per particle $E/N$ (see
the green dot-dashed line). For large trapping frequency, such a jump
disappears (see Fig. \ref{fig13_CMTrap}(b)). Instead, we start to
observe the formation of a dip structure in the breathing mode frequency.
The jump or discontinuity at small trapping frequency is easy to understand.
It is simply the first-order transition from the droplet state to
the gas-like state that we already discussed in Fig. \ref{fig2_NcNm}
in the absence of the external harmonic trap. We should also find
a metastable state for a small window in the number of particles,
if we try different Gaussian ansatz (with different width) as the
initial state for solving the stationary GPE and then the Bogoliubov
equations. Indeed, as shown in Fig. \ref{fig15_PhaseDiagram}, we
can determine the critical and threshold numbers of particles, $N_{c}$
and $N_{m}$, at different trapping potential. Remarkably, the window
for the metastable state shrinks with increasing trapping frequency
$\omega_{0}$. It closes completely at the tri-critical point $\omega_{0}\sim0.032$
(see, i.e., the orange circle in Fig. \ref{fig15_PhaseDiagram}).
Above this value, the transition from the droplet state to the gas-like
phase becomes smooth (i.e., second-order) and we mark the dip position
in the breathing mode frequency as the transition point.

Finally, it can be readily seen from Fig. \ref{fig13_CMTrap} that,
although the excitation spectrum changes a lot under the external
trapping potential, the qualitative behavior of the surface mode frequencies,
as a function of the number of particles, turn out to be very robust.
They decrease with increasing number of particles, following the same
pattern as in the absence of the external trapping potential (i.e..
compared to Fig. \ref{fig6_CollectiveModes}(a)). To better understand
this, let us now check more carefully the quasi-particle wave-functions
of the surface modes.

\begin{figure}[t]
\begin{centering}
\includegraphics[width=0.48\textwidth]{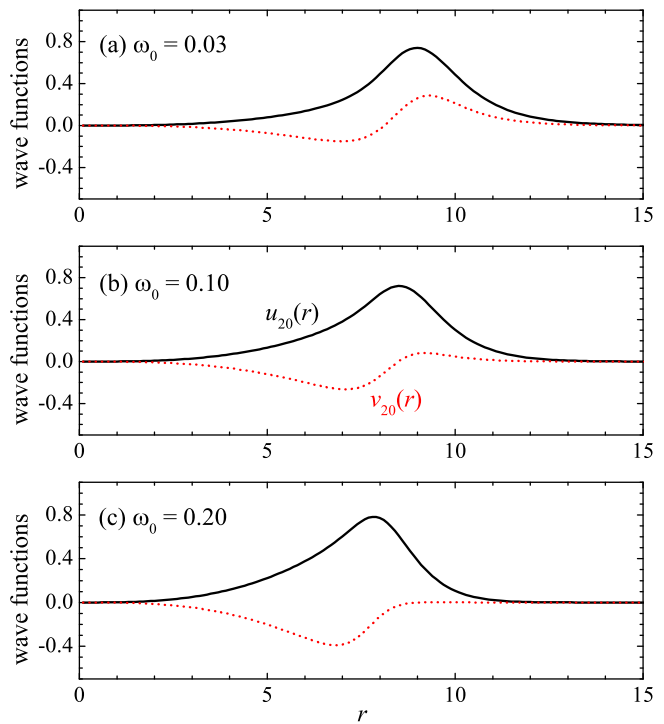}
\par\end{centering}
\caption{\label{fig16_uvLargeNTrap} The quasi-particle wave-functions of the
lowest surface mode $\omega_{20}$ at three trapping frequencies $\omega_{0}=0.03$
(a), $\omega_{0}=0.10$ (b) and $\omega_{0}=0.20$ (c) and at a large
reduced particle number $N=3000$. The solid and dashed lines show
$u_{20}(r)$ and $v_{20}(r)$, respectively. Here, we take the parameter
$\gamma=1/2$.}
\end{figure}

\subsection{Surface modes}

In Fig. \ref{fig16_uvLargeNTrap}, we show the wave-functions $u_{20}(r)$
(solid lines) and $v_{20}(r)$ (dashed lines) of the lowest surface
mode at $N=3000$ and at three different trapping potentials. We find
that the shapes of the wave-functions are qualitatively unchanged
with increasing trapping frequency, although there is a slight shift
in the peak position in $u_{20}(r)$, This shift is presumably due
to the reduced droplet size, since the external trapping potential
provides additional confinement to the particles. Indeed, by examining
the density profiles in Fig. \ref{fig11_DensityTrap}, we observe
that the droplet edge shifts from $R\sim9$ to $R\sim8.5$, and finally
to $R\sim8$, when we increase the trapping frequency from $\omega_{0}=0.03$
to $\omega_{0}=0.10$, and to $\omega_{0}=0.20$, which is consistent
with the peak position found in the wave-functions $u_{20}(r)$.

\begin{figure}[t]
\begin{centering}
\includegraphics[width=0.48\textwidth]{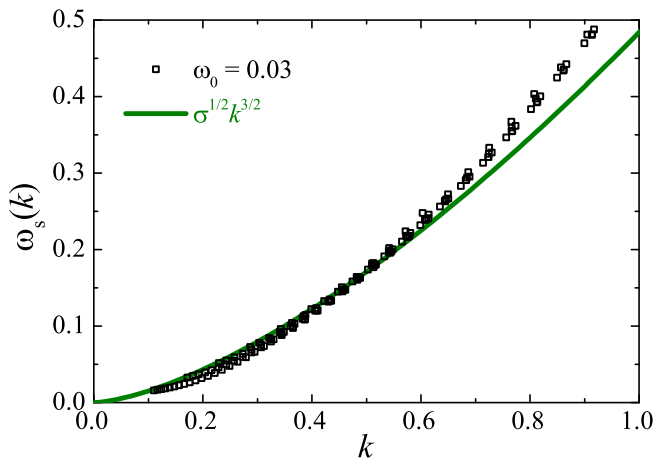}
\par\end{centering}
\caption{\label{fig17_wk15Trap} Excitation frequencies of the surface modes
$\omega_{l,n=0}$ from $l=2$ to $l=9$, for some selected reduced
particle numbers $N$ ranging from $500$ to $30000$, as a function
of the effective wave-vector $k=[l(l-1)(l+2)]^{1/3}/R$ at the harmonic
trapping potential $\omega_{0}=0.03$ (squares). As the external trapping
potential gives rise to a background contribution ($\sim\omega_{0}$)
to the mode frequency, we have defined $\omega_{s}\simeq\omega_{l0}-\omega_{0}$.
The green thick line shows the anticipated dispersion relation $\omega_{s}(k)=\sqrt{\sigma_{s}}k^{3/2}$,
where the dimensionless surface tension $\sigma_{s}=6(1+\sqrt{3})/35\simeq0.234176$
at the parameter $\gamma=1/2$.}
\end{figure}

It is now natural to ask, could we experimentally confirm the exotic
$k^{3/2}$ dispersion relation for the surface modes under a reasonably
small external trapping potential? We consider the strategy adopted
earlier and use Eq. (\ref{eq:Radius}) to convert the angular momentum
$2\leq l\leq9$ to an effective wave-vector $k$ at different number
of particles $N\subseteq(500,30000$). In this interval, we find that
the surface mode frequency is typically pushed up by the external
trapping potential by an amount $\sim\omega_{0}$. Therefore, we subtract
this background contribution and define $\omega_{s}\simeq\omega_{l0}-\omega_{0}$.
In Fig. \ref{fig17_wk15Trap}, we show $\omega_{s}$ as a function
of the effective wave-vector $k$ at a weak trapping frequency $\omega_{0}=0.03$.
We observe that, overall the data points roughly follow the expected
dispersion relation $\omega_{s}(k)=\sqrt{\sigma_{s}}k^{3/2}$, which
is shown by a thick green line. However, we can not find a perfect
data collapse at large number of particles as in the self-bound droplet
(cf. Fig. \ref{fig10_wk15}). The data points become more scattered
as we increase the trapping potential to $\omega_{0}=0.20$ (not shown
in the figure). Presumably, this is due to the difficulty in determining
the radius $R$ of the droplet: when the external trapping frequency
increases, the flat-top structure in the density profile ceases to
exist and the sharp edge becomes less well-defined. The inaccurate
determination of the droplet radius in turn makes the effective wave-vector
$k$ ill-defined and hence leads to scattered data points in the dispersion
relation. Nevertheless, at the weak external trapping potential considered
in Fig. 17, the scattering of the data points is not obvious and the
nonlinearity of the dispersion is evident. We may then measure the
surface tension from a curve fitting to the anticipated $k^{3/2}$
dispersion relation at small momentum.

\begin{figure}[t]
\begin{centering}
\includegraphics[width=0.48\textwidth]{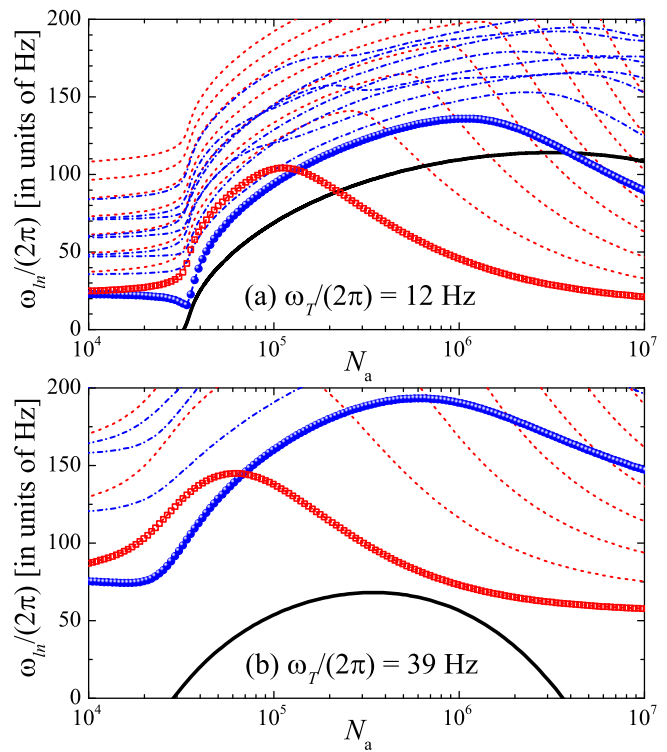}
\par\end{centering}
\caption{\label{fig18_exptCMTrap} Excitation frequencies $\omega_{ln}$ ($l\protect\leq9$
and $n\protect\leq2$) of a $^{39}$K binary mixture at the magnetic
field $B=56.453$ G, as a function of the actual particle number $N_{a}$
at the trapping frequency $\omega_{T}=2\pi\times12$ Hz (a, upper
panel) and $\omega_{T}=2\pi\times39$ Hz (b, lower panel). The red
dashed lines show the surface modes $\omega_{l\protect\geq2,n=0}$
and the blue dot-dashed lines shows the other bulk modes. The lowest
surface mode $\omega_{20}$ (i.e, quadruple mode) and the lowest bulk
mode (breathing monopole mode) are shown by the red open squares and
blue solid circles, respectively. The black thick lines show $-\mu_{a}$.
Here, the parameter $\gamma=0.373<1/2$ according to the parameterization
to the DMC equation of state \citep{Cikojevic2020arXiv}. We note
that, the breathing mode ($\omega_{00}$) and quadrupole mode ($\omega_{20}$)
without trapping potential have recently been investigated by Cikojevi\'{c}
and co-workers \citep{Cikojevic2020arXiv}.}
\end{figure}

\begin{figure}[t]
\begin{centering}
\includegraphics[width=0.48\textwidth]{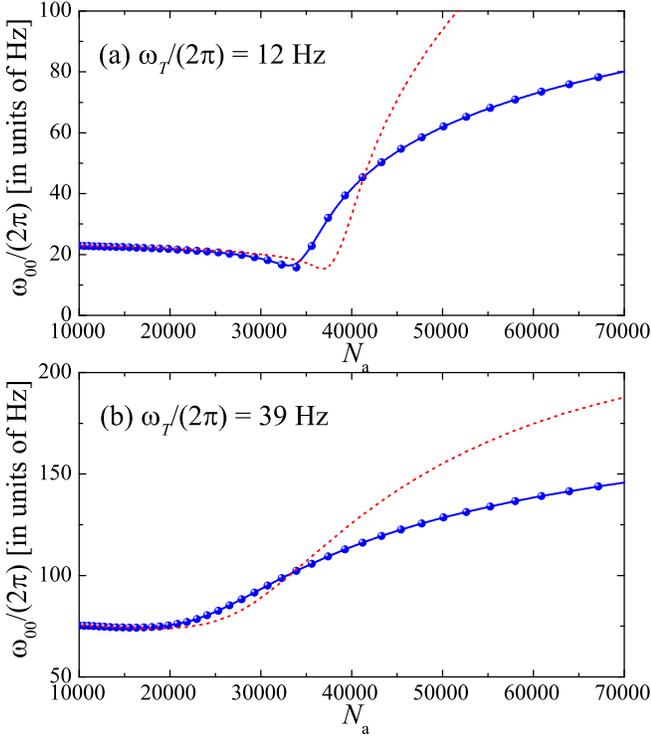}
\par\end{centering}
\caption{\label{fig19_GaussianWBTrap} Breathing mode frequencies $\omega_{00}$
of a $^{39}$K binary mixture at the magnetic field $B=56.453$ G,
as a function of the actual particle number $N_{a}$ near the gas-liquid
transition at the trapping frequency $\omega_{T}=2\pi\times12$ Hz
(a, upper panel) and $\omega_{T}=2\pi\times39$ Hz (b, lower panel).
The solid lines with blue circles show the results of the linearized
Bogoliubov equations, while the red dashed lines show the predictions
from the Gaussian ansatz.}
\end{figure}

\section{Experimental relevances}

To make connection with the experiments, let us focus on the binary
$^{39}$K mixtures \citep{Cabrera2018,Semeghini2018}. In this case,
an accurate calculation of the total energy functional $E(n)$ is
difficult, since the intra- and inter-species inter-particle interactions
both involve a large effective range of interactions. Here, we consider
the recent equation of state obtained by DMC and the related parameterization
\citep{Cikojevic2020,Cikojevic2020arXiv}. Let us choose a typical
magnetic field $B=56.453$ G, at which the parameters in the total
energy functional $E(n)=-A_{0}n^{2}+A_{1}n^{2+\gamma}$ are given
by \citep{Cikojevic2020arXiv},
\begin{eqnarray}
A_{0} & = & \alpha\frac{\hbar^{2}a_{11}}{2m},\\
A_{1} & = & \beta\frac{\hbar^{2}a_{11}^{3\gamma+1}}{2m},
\end{eqnarray}
with\emph{ }$\alpha=0.423$, $\beta=8.550$, $\gamma=0.373$ and $a_{11}=70.119a_{\textrm{bohr}}\simeq3.711$
nm. By using Eq. (\ref{eq:N0}) and Eq. (\ref{eq:EnergyUnits}), it
is straightforward to obtain,
\begin{eqnarray}
\frac{\hbar^{2}}{m\xi^{2}} & = & \frac{1}{2\beta^{\frac{1}{\gamma}}}\left[\frac{\alpha}{\left(1+\gamma\right)}\right]^{\frac{1+\gamma}{\gamma}}\frac{\hbar^{2}}{ma_{11}^{2}}\simeq2\pi\hbar\times392\textrm{ Hz},\\
n_{0}\xi^{3} & = & \sqrt{8}\beta^{\frac{1}{2\gamma}}\left[\frac{\alpha}{\left(1+\gamma\right)}\right]^{\frac{1+3\gamma}{2\gamma}}\simeq1423.
\end{eqnarray}
Therefore, the residual harmonic trapping frequency $\omega_{T}\sim2\pi\times12$
Hz in the LENS experiment \citep{Semeghini2018} translates into the
dimensionless trapping frequency 
\begin{equation}
\omega_{0}=\frac{\hbar\omega_{T}}{\hbar^{2}/\left(m\xi^{2}\right)}=\frac{12}{392}\simeq0.03,
\end{equation}
and the maximum actual number of particles $N_{a}\sim4\times10^{5}$
would correspond to a reduced number of particles 
\begin{equation}
N=\frac{N_{a}}{n_{0}\xi^{3}}=\frac{4\times10^{5}}{1423}\simeq281,
\end{equation}
which is already large enough to support a discrete surface mode with
$l=2$.

In Fig. \ref{fig18_exptCMTrap}, we show the excitation spectrum of
a binary $^{39}$K mixture at $B=56.453$ G, as a function of the
actual number of particles $N_{a}$, at the two characteristic trapping
frequencies $\omega_{T}=2\pi\times12$ Hz (a) and $\omega_{T}=2\pi\times39$
Hz (b), which correspond to $\omega_{0}=0.03$ and $\omega_{0}=0.10$,
respectively. Experimentally, it would be interesting to confirm the
non-monotonic dependence of the $l=2$ surface mode frequency on the
number of particles $N_{a}$, as illustrated by the red squares. For
the case of a small trapping frequency $\omega_{T}=2\pi\times12$
Hz in (a), the peak position might be interpreted as $N_{\textrm{th}}$
($\times n_{0}\xi^{3}$). We may also qualitatively determine the
surface tension $\sigma_{s}$ by fitting the surface mode frequency
at the large number of particles with Eq. (\ref{eq:wsurface}), after
the background contribution $\sim\omega_{T}$ being subtracted.

To close the section, let us briefly comment on the applicability
of the Gaussian variational approach for a three-dimensional quantum
droplet under the external trapping potential, which has been used
in the previous studies \citep{Cappellaro2018,Cappellaro2017}. As
can be seen from Fig. \ref{fig19_GaussianWBTrap}, the Gaussian variational
approach predicts very accurate breathing mode frequency in the gas-like
state. However, it strongly over-estimates the mode frequency in the
droplet state. For the description of the gas-droplet transition,
it works qualitatively well and over-estimates the critical number
of particles by several tens of percent.

\section{Conclusions}

In summary, we have presented a systematic investigation of collective
excitations of a three-dimensional ultradilute Bose droplet, by using
a phenomenological low-energy effective theory \citep{Petrov2015}.
Two approaches have been considered: one is the approximate Gaussian
variational approach and another is the numerically exact solution
of the Bogoliubov equations. An energy density functional $E(n)$
with a general exponent $\gamma$, i.e., $E(n)=-A_{0}n^{2}+A_{1}n^{2+\gamma}$,
has been adopted to provide an accurate equation of state for the
droplet \citep{Cikojevic2019,Cikojevic2020}. Furthermore, we have
considered the effect of a small external harmonic trapping potential,
which may be experimentally used to enhance the stability of the system.

We have found that the first-order droplet-to-gas transition and the
excitation spectrum can sensitively depend on the parameter $\gamma$.
Yet, in the absence of the external trapping potential, the structure
of the spectrum is qualitatively unchanged and is universal. It consists
of the discrete bulk modes and surface modes, below the particle-emission
threshold $-\mu>0$. For sufficiently large number of particles, they
assume the well-known dispersion relations $\omega=ck$ and $\omega_{s}=\sqrt{\sigma_{s}}k^{3/2}$,
respectively, if we properly define the effective wave-vector $k$.

In the presence of a weak external harmonic trap, we have observed
that the droplet-to-gas transition and the excitation spectrum also
change significantly. The phase window for the metastable state shrinks
quickly with increasing external trapping potential. Above a tri-critical
trapping potential (i.e., $\omega_{0}\simeq0.032$ at $\gamma=1/2$),
the droplet-to-gas transition becomes smooth. Despite the significant
change in the excitation spectrum due to the external trapping potential,
we have found that the qualitative behavior of the surface modes are
very robust. We have shown that it is possible to experimentally confirm
their exotic $k^{3/2}$ dispersion relation.

We have also calculated the excitation spectrum for a binary $^{39}$K
mixture at a typical magnetic field, with the help of the accurate
equation of state recently obtained from the diffusion Monte Carlo
simulations \citep{Cikojevic2020arXiv}. A non-monotonic dependence
of the $l=2$ surface mode frequency on the actual number of particles
has been predicted, under realistic trapping potentials. Both the
peak structure and the decrease in the surface mode frequency at large
number of particles should be observable in the current experimental
configuration. We note, however, that experimentally the three-body
loss could become severe at large number of particles. In this respect,
a heteronuclear $^{41}$K-$^{87}$Rb mixture with much longer lifetime
\citep{DErrico2019} could be a better candidate to experimentally
confirm our predictions on collective excitations. This possibility
will be explored in a future study.
\begin{acknowledgments}
We thank Zhichao Guo and Dajun Wang for stimulating discussions. This
research was supported by the Australian Research Council's (ARC)
Discovery Program, Grant No. DP170104008 (H.H.) and Grant No. DP180102018
(X.-J.L).
\end{acknowledgments}


\begin{thebibliography}{10}
\bibitem{Petrov2015}D. S. Petrov, Quantum Mechanical Stabilization
of a Collapsing Bose-Bose Mixture, Phys. Rev. Lett. \textbf{115},
155302 (2015).

\bibitem{FerrierBarbut2016}I. Ferrier-Barbut, H. Kadau, M. Schmitt,
M. Wenzel, and T. Pfau, Observation of Quantum Droplets in a Strongly
Dipolar Bose Gas, Phys. Rev. Lett. \textbf{116}, 215301 (2016).

\bibitem{Schmitt2016}M. Schmitt, M. Wenzel, F. Böttcher, I. Ferrier-Barbut,
and T. Pfau, Self-bound droplets of a dilute magnetic quantum liquid,
Nature (London) \textbf{539}, 259 (2016).

\bibitem{Chomaz2016}L. Chomaz, S. Baier, D. Petter, M. J. Mark, F.
Wächtler, L. Santos, and F. Ferlaino, Quantum-Fluctuation-Driven Crossover
from a Dilute Bose-Einstein Condensate to a Macrodroplet in a Dipolar
Quantum Fluid, Phys. Rev. X \textbf{6}, 041039 (2016).

\bibitem{Cabrera2018}C. Cabrera, L. Tanzi, J. Sanz, B. Naylor, P.
Thomas, P. Cheiney, and L. Tarruell, Quantum liquid droplets in a
mixture of Bose-Einstein condensates, Science \textbf{359}, 301 (2018).

\bibitem{Cheiney2018}P. Cheiney, C. R. Cabrera, J. Sanz, B. Naylor,
L. Tanzi, and L. Tarruell, Bright Soliton to Quantum Droplet Transition
in a Mixture of Bose-Einstein Condensates, Phys. Rev. Lett. \textbf{120},
135301 (2018).

\bibitem{Semeghini2018}G. Semeghini, G. Ferioli, L. Masi, C. Mazzinghi,
L. Wolswijk, F. Minardi, M. Modugno, G. Modugno, M. Inguscio, and
M. Fattori, Self-Bound Quantum Droplets of Atomic Mixtures in Free
Space, Phys. Rev. Lett. \textbf{120}, 235301 (2018).

\bibitem{Ferioli2019}G. Ferioli, G. Semeghini, L. Masi, G. Giusti,
G. Modugno, M. Inguscio, A. Gallemi, A. Recati, and M. Fattori, Collisions
of Self-Bound Quantum Droplets, Phys. Rev. Lett. \textbf{122}, 090401
(2019).

\bibitem{Tanzi2019PRL}L. Tanzi, E. Lucioni, F. Famà, J. Catani, A.
Fioretti, C. Gabbanini, R.\LyXThinSpace N. Bisset, L. Santos, and
G. Modugno, Observation of a Dipolar Quantum Gas with Metastable Supersolid
Properties, Phys. Rev. Lett. \textbf{122}, 130405 (2019).

\bibitem{DErrico2019}C. D\textquoteright Errico, A. Burchianti, M.
Prevedelli, L. Salasnich, F. Ancilotto, M. Modugno, F. Minardi, and
C. Fort, Observation of quantum droplets in a heteronuclear bosonic
mixture, Phys. Rev. Research \textbf{1}, 033155 (2019).

\bibitem{Bottcher2019}F. Böttcher, M. Wenzel, J.-N. Schmidt, M. Guo,
T. Langen, I. Ferrier-Barbut, T. Pfau, R. Bombín, J. Sánchez-Baena,
J. Boronat, and F. Mazzanti, Dilute dipolar quantum droplets beyond
the extended Gross-Pitaevskii equation, Phys. Rev. Research \textbf{1},
033088 (2019).

\bibitem{Bottcher2020}For a recent review, see, for example, F. Böttcher,
J.-N. Schmidt, J. Hertkorn, K. S. H. Ng, S. D. Graham, M. Guo, T.
Langen, and T. Pfau, New states of matter with fine-tuned interactions:
quantum droplets and dipolar supersolids, arXiv:2007.06391 (2020). 

\bibitem{LeeHuangYang1957}T. D. Lee, K. Huang, and C. N. Yang, Eigenvalues
and Eigenfunctions of a Bose System of Hard Spheres and Its Low-Temperature
Properties, Phys. Rev. \textbf{106}, 1135 (1957).

\bibitem{Chin1995}S. A. Chin and E. Krotscheck, Surface Excitations
of Helium Droplets, Phys. Rev. Lett. \textbf{74}, 1143 (1995).

\bibitem{Dalfovo1999}F. Dalfovo, S. Giorgini, L. P. Pitaevskii, and
S. Stringari, Theory of Bose-Einstein condensation in trapped gases,
Rev. Mod. Phys. \textbf{71}, 463 (1999).

\bibitem{Kinast2004}J. Kinast, S. L. Hemmer, M. E. Gehm, A. Turlapov,
and J. E. Thomas, Evidence for Superfluidity in a Resonantly Interacting
Fermi Gas, Phys. Rev. Lett. \textbf{92}, 150402 (2004).

\bibitem{Bartenstein2004}M. Bartenstein, A. Altmeyer, S. Riedl, S.
Jochim, C. Chin, J. Hecker Denschlag, and R. Grimm, Collective Excitations
of a Degenerate Gas at the BEC-BCS Crossover, Phys. Rev. Lett. \textbf{92},
203201 (2004).

\bibitem{Hu2004}H. Hu, A. Minguzzi, X.-J. Liu, and M. P. Tosi, Collective
Modes and Ballistic Expansion of a Fermi Gas in the BCS-BEC Crossover,
Phys. Rev. Lett. \textbf{93}, 190403 (2004).

\bibitem{Holten2018}M. Holten, L. Bayha, A. C. Klein, P. A. Murthy,
P. M. Preiss, and S. Jochim, Anomalous Breaking of Scale Invariance
in a Two-Dimensional Fermi Gas, Phys. Rev. Lett. \textbf{121}, 120401
(2018).

\bibitem{Peppler2018}T. Peppler, P. Dyke, M. Zamorano, S. Hoinka,
and C. J. Vale, Quantum Anomaly and 2D-3D Crossover in Strongly Interacting
Fermi Gases, Phys. Rev. Lett. \textbf{121}, 120402 (2018).

\bibitem{Hu2019}H. Hu, B. C. Mulkerin, U. Toniolo, L. He, and X.-J.
Liu, Reduced Quantum Anomaly in a Quasi-Two-Dimensional Fermi Superfluid:
Significance of the Confinement-Induced Effective Range of Interactions,
Phys. Rev. Lett. \textbf{122}, 070401 (2019).

\bibitem{Yin2020}X.\LyXThinSpace Y. Yin, H. Hu, and X.-J. Liu, Few-Body
Perspective of a Quantum Anomaly in Two-Dimensional Fermi Gases, Phys.
Rev. Lett. \textbf{124}, 013401 (2020).

\bibitem{Tanzi2019Nature}L. Tanzi, S. M. Roccuzzo, E. Lucioni, F.
Famà, A. Fioretti, C. Gabbanini, G. Modugno, A. Recati, and S. Stringari,
Supersolid symmetry breaking from compressional oscillations in a
dipolar quantum gas, Nature (London) \textbf{574}, 382 (2019).

\bibitem{Guo2019}M. Guo, F. Böttcher, J. Hertkorn, J.-N. Schmidt,
M. Wenzel, H. P. Büchler, T. Langen, and T. Pfau, The low-energy Goldstone
mode in a trapped dipolar supersolid, Nature (London) \textbf{574},
386 (2019).

\bibitem{Baillie2016}D. Baillie, R. M. Wilson, R. N. Bisset, and
P. B. Blakie, Self-bound dipolar droplet: A localized matter wave
in free space, Phys. Rev. A \textbf{94}, 021602(R) (2016).

\bibitem{Wachtler2016}F. Wächtler and L. Santos, Ground-state properties
and elementary excitations of quantum droplets in dipolar Bose-Einstein
condensates, Phys. Rev. A \textbf{94}, 043618 (2016).

\bibitem{Baillie2017}D. Baillie, R.\LyXThinSpace M. Wilson, and P.\LyXThinSpace B.
Blakie, Collective Excitations of Self-Bound Droplets of a Dipolar
Quantum Fluid, Phys. Rev. Lett. \textbf{119}, 255302 (2017).

\bibitem{NoteNumberNodes}For nonzero angular momentum $l\neq0$,
the radial quantum number $n$ denotes the number of nodes in the
radial wavefunctions. However, at zero angular momentum, the nodeless
wavefunction is the condensate wave-function and is excluded as a
wave-function of Bogoliubov quasi-particles. In the $l=0$ sector,
therefore, the number of nodes in the radial wavefunctions is given
by $n+1$. Hence, the breathing mode has a node in its radial wavefunction. 

\bibitem{Cikojevic2019}V. Cikojevi\'{c}, L. Vranješ Markic, G. E.
Astrakharchik, and J. Boronat, Universality in ultradilute liquid
Bose-Bose mixtures, Phys. Rev. A \textbf{99}, 023618 (2019).

\bibitem{Cikojevic2020}V. Cikojevi\'{c}, L. Vranješ Marki\'{c}, and
J. Boronat, Finite-range effects in ultradilute quantum drops, New
J. Phys. \textbf{22}, 053045 (2020).

\bibitem{Cikojevic2020arXiv}V. Cikojevi\'{c}, L. Vranješ Marki\'{c},
M. Pi, M. Barranco, and J. Boronat, Towards a QMC-based density functional
including finite-range effects: Excitation modes of a $^{39}$K quantum
droplet, Phys. Rev. A \textbf{102}, 033335 (2020).

\bibitem{Bulgac2002}A. Bulgac, Dilute Quantum Droplets, Phys. Rev.
Lett. \textbf{89}, 050402 (2002).

\bibitem{Blume2002}D. Blume, B. D. Esry, Chris H. Greene, N. N. Klausen,
and G. J. Hanna, Formation of Atomic Tritium Clusters and Bose-Einstein
Condensates, Phys. Rev. Lett. \textbf{89}, 163402 (2002).

\bibitem{Mestrom2020}P.\LyXThinSpace M.\LyXThinSpace A. Mestrom,
V.\LyXThinSpace E. Colussi, T. Secker, G.\LyXThinSpace P. Groeneveld,
and S.\LyXThinSpace J.\LyXThinSpace J.\LyXThinSpace M.\LyXThinSpace F.
Kokkelmans, van der Waals Universality near a Quantum Tricritical
Point, Phys. Rev. Lett. \textbf{124}, 143401 (2020).

\bibitem{Stringari1987}S. Stringari and J. Treiner, Surface properties
of liquid $^{3}$He and $^{4}$He: A density-functional approach,
Phys. Rev. B \textbf{36}, 8369 (1987).

\bibitem{Ferioli2020}G. Ferioli, G. Semeghini, S. Terradas-Briansó,
L. Masi, M. Fattori, and M. Modugno, Dynamical formation of quantum
droplets in a $^{39}$K mixture, Phys. Rev. Research \textbf{2}, 013269
(2020).

\bibitem{Cappellaro2018}A. Cappellaro, T. Macrì, and L. Salasnich,
Collective modes across the soliton-droplet crossover in binary Bose
mixtures, Phys. Rev. A \textbf{97}, 053623 (2018).

\bibitem{Astrakharchik2018}G. E. Astrakharchik and B. A. Malomed,
Dynamics of one-dimensional quantum droplets, Phys. Rev. A \textbf{98},
013631 (2018).

\bibitem{Tylutki2020}M. Tylutki, G. E. Astrakharchik, B. A. Malomed,
and D. S. Petrov, Collective excitations of a one-dimensional quantum
droplet, Phys. Rev. A \textbf{101}, 051601(R) (2020).

\bibitem{Hu2020a}H. Hu and X.-J. Liu, Consistent theory of self-bound
quantum droplets with bosonic pairing, arXiv:2005.08581v2 (2020);
to appear in Physical Review Letters.

\bibitem{Hu2020b}H. Hu, J. Wang, and X.-J. Liu, Microscopic pairing
theory of a binary Bose mixture with interspecies attractions: Bosonic
BEC-BCS crossover and ultradilute low-dimensional quantum droplets,
Phys. Rev. A \textbf{102}, 043301 (2020).

\bibitem{Hu2020c}H. Hu and X.-J. Liu, Microscopic derivation of the
extended Gross-Pitaevskii equation for quantum droplets in binary
Bose mixtures, Phys. Rev. A \textbf{102}, 043302 (2020).

\bibitem{Cappellaro2017}A. Cappellaro, T. Macrì, G. F. Bertacco,
and L. Salasnich, Equation of state and self-bound droplet in Rabi-coupled
Bose mixtures, Sci. Rep. \textbf{7}, 13358 (2017).

\bibitem{Menotti2002}C. Menotti and S. Stringari, Collective oscillations
of a one-dimensional trapped Bose-Einstein gas, Phys. Rev. A \textbf{66},
043610 (2002).

\bibitem{Hu2014}H. Hu, G. Xianlong, and X.-J. Liu, Collective modes
of a one-dimensional trapped atomic Bose gas at finite temperatures,
Phys. Rev. A \textbf{90}, 013622 (2014).

\bibitem{Pu1998}H. Pu and N. P. Bigelow, Properties of Two-Species
Bose Condensates, Phys. Rev. Lett. \textbf{80}, 1130 (1998).

\bibitem{Hutchinson1997}D. A. W. Hutchinson, E. Zaremba, and A. Griffin,
Finite Temperature Excitations of a Trapped Bose Gas, Phys. Rev. Lett.
\textbf{78}, 1842 (1997).

\bibitem{FluidMechanicsBook1987}L. D. Landau and E. M. Lifshitz,
\textit{Fluid Mechanics} (Pergamon Press, Oxford, 1987), $\mathsection61$
and $\mathsection62$. 
\end{thebibliography}
\end{document}